\documentclass[reprint,aps,twocolumn]{revtex4-2}
\usepackage{amsmath}
\usepackage{amssymb}
\usepackage{graphicx}
\usepackage{subfigure}
\usepackage{algorithm}
\usepackage{bm}
\usepackage{color}

\usepackage{commath}
\usepackage{dcolumn}
\usepackage{multirow}

\usepackage{hyperref}

\begin{document}

\preprint{ITP307/2022-03}

\title{Energy--Information Trade-off Induces Continuous and Discontinuous Phase Transitions in Lateral Predictive Coding}

\author{Zhen-Ye Huang$^{1,2,*}$, Ruyi Zhou$^{3,*}$, Miao Huang$^{1,2,*}$, and Hai-Jun Zhou$^{1,2,4}$}

\affiliation{
  $^1$CAS Key Laboratory for Theoretical Physics, Institute of Theoretical Physics, Chinese Academy of Sciences, Beijing 100190, China\\
  $^2$School of Physical Sciences, University of Chinese Academy of Sciences, Beijing 100049, China\\
  $^3$School of Optics and Photonics, Beijing Institute of Technology, Beijing 100081, China\\
  $^4$MinJiang Collaborative Center for Theoretical Physics, MinJiang University, Fuzhou 350108, China \\
  $^*$These authors (ZYH, RZ, MH) contributed equally to this work.
}

\date{\today}

\begin{abstract}
  Lateral predictive coding is a recurrent neural network which creates energy-efficient internal representations by exploiting statistical regularity in sensory inputs. Here we investigate the trade-off between information robustness and energy in a linear model of lateral predictive coding analytically and by numerical minimization of a free energy. We observe several phase transitions in the synaptic weight matrix, especially a continuous transition which breaks reciprocity and permutation symmetry and builds cyclic dominance and a discontinuous transition with the associated sudden emergence of tight balance between excitatory and inhibitory interactions. The optimal network follows an ideal-gas law in an extended temperature range and saturates the efficiency upper-bound of energy utilization. These results bring theoretical insights on the emergence and evolution of complex internal models in predictive processing systems.
\end{abstract}

\maketitle


\subsection{Introduction}

Predictive coding is an influential theory in computational neuroscience with a long history tracing back to Kant and Helmhotz~\cite{Swanson-2016,Wade-2021,Barlow-1961,Huang-Rao-2011}, and its recent conceptual advances include the Bayesian brain theory and the free energy principle of information processing and decision making~\cite{Kim-2021,Friston-etal-2006,Aguilera-etal-2022,Jirsa-Sheheitli-2022}. The predictive coding theory regards the brain as a multi-layered hierarchical neuron network which builds an internal model for the external world and employs it to perceive sensory signals and make predictions. So far most studies on predictive coding have focused on the feedforward and feedback interactions between different layers, and have largely ignored lateral (horizontal, recurrent) interactions within the same layer of neurons. But lateral interactions are abundant in biological neural networks and they play significant roles in perception and inference~\cite{Srinivasan-etal-1982,Tang-etal-2018,Pang-etal-2021,Fumarola-etal-2022}. Children brain consumes $50\%$ of the body's total metabolic energy and this fraction decreases to $20\%$ in adulthood~\cite{Clarke-Sokoloff-1999}. As a big energy consumer it is vital for the brain to be capable of achieving energy-efficient information processing~\cite{Lennie-2003,Chen-etal-2013,Niven-2016,Yu-Yu-2017,Metzner-etal-2023}. Lateral predictive coding (LPC) might have been an elegant evolutionary solution to cope with the brain's tight energy budget.

One basic function of lateral interactions is to reduce the energy cost of internal representations.  By adapting the synaptic weights according to the input correlation between spatially adjacent neurons, the external drive to one neuron is partially cancelled by the messages from its neighboring neurons and its activity becomes weaker and sparser~\cite{Jutten-Herault-1991,Olshausen-Field-1996,Harpur-Prager-1996,Ali-etal-2022}. On the other hand, reducing the magnitude and redundancy of internal representation compromises its information content, making it less robust to external and internal noises~\cite{Yu-Liu-2014,Padamsey-etal-2022,Weninger-etal-2022}. How to balance the conflicting demands of energy reduction and information robustness is an important issue of LPC. Yet LPC has been a rarely discussed topic in the physics community, in stark contrast to the huge and enduring theoretical enthusiasm on the other basic neural network models such as the perceptron, the Hopfield model of associative memory and the restricted Boltzmann machine~\cite{Huang-2022}.

Here we study phase transitions induced by energy--information trade-off in the simplest model of linear LPC~\cite{Huang-etal-2022}. We introduce a temperature parameter $T$ to incorporate the fitness effect of information robustness and define a free energy, which binds together energy and function, as the minimization objective~\cite{Bell-Sejnowski-1995,Tishby-etal-1999,He-Wang-2017}. We find that even if the input signals are statistically symmetric among all the $N$ units, the optimal synaptic weight matrix will spontaneously break permutation (reciprocal) symmetry at certain critical temperature and form cyclic-dominant interaction patterns. An ideal-gas law $E = (N/2)T$ is then followed by the energy $E$ of the multi-unit ($N\geq 3$) LPC networks, achieving upper-bound efficiency in energy utilization. At low temperatures the weight matrix experiences several further phase transitions, one of them being a discontinuous transition to tight balance of excitatory and inhibitory interactions. Internal symmetries within sub-groups of units may also break, leading to multi-level functional differentiation.

Our theoretical results reveal the emergence of structural complexity in LPC systems. These collective properties of non-reciprocity, correlation reduction, and excitation--inhibition (EI) balance are qualitatively similar to the empirical observations on real-life recurrent neural networks. As an initial attempt we have left many important issues untouched, such as nonlinear synaptic interactions and non-quadratic energetic cost. More efforts from the physics community are needed to fully appreciate LPC as a basic neural computing circuit.

\subsection{Model and free energy}

Consider a fully connected network of $N$ units (each may be a single neuron or a small region of the brain). Denote an internal state vector as $\vec{\bm{x}} = (x_1, \ldots, x_N)^\top$ and a sensory input as $\vec{\bm{s}} = (s_1, \ldots, s_N)^\top$. Upon receiving an input $\vec{\bm{s}}$, unit $i$ responds by changing its state $x_i$ according to a linear LPC dynamics~\cite{Jutten-Herault-1991,Harpur-Prager-1996,Huang-etal-2022},
\begin{equation}
  \frac{{\rm d} x_i}{ {\rm d} t} =
  s_i - x_i - \sum\limits_{j \neq i} w_{i j} x_j
  \; ,
  \label{eq:pcr}
\end{equation}
where positive (negative) synaptic weight $w_{i j}$ means that unit $j$ inhibits (excites) unit $i$. The steady-state encoding from sensory input to internal representation is $\vec{\bm{x}}_{\vec{\bm{s}}} = (\bm{I} + \bm{W})^{-1} \vec{\bm{s}}$ with $\bm{I}$ being the identity matrix and $\bm{W}$ the synaptic weight matrix. The convergence of the dynamics (\ref{eq:pcr}) requires the real part of every eigenvalue of $(\bm{I}+\bm{W})$ be positive (we set a threshold value $10^{-5}$ and call it the eigenvalue bottomline). We interpret $\sum_{j\neq i} w_{i j} x_j$ as the predicted sensory input to unit $i$, and $x_i$ as the prediction error between the actual $s_i$ and the predicted value~\cite{Srinivasan-etal-1982}. Assuming the energy cost of maintaining and transmitting a prediction error be quadratic in $x_i$, the mean energy $E$ of an internal state is
\begin{equation}
  E \equiv \sum_{\vec{\bm{s}}}
  {\rm P}(\vec{\bm{s}}) {\vec{\bm{x}}}_{\vec{\bm{s}}}^2
  = \textrm{Tr}\Bigl[ \bigl(\bm{I}+\bm{W}\bigr)^{-1}
    \bm{C} \bigl(\bm{I}+\bm{W}^\top\bigr)^{-1} \Bigr]
  \; ,
\end{equation}
where ${\rm P}(\vec{\bm{s}})$ is the probability distribution of sensory inputs and $\bm{C}$ is the input correlation matrix with elements $c_{i j} \equiv \sum_{\vec{\bm{s}}} {\rm P}(\vec{\bm{s}}) s_i s_j$. The total energy cost of representing $\mathcal{M} \gg 1$ sensory inputs is then $\mathcal{M} E$.

An infinitesimal volume of the sensory space is transformed by $\vec{\bm{x}}_{\vec{\bm{s}}}$ to an infinitesimal volume in the internal representation space, with Jacobian $1/{\rm Det}(\bm{I} + \bm{W})$ which is the inverse determinant. The entropy of the internal states $\bm{x}$ is then
\begin{equation}
  S = - \log \bigl[ {\rm Det}(\bm{I} +  \bm{W}) \bigr]
\end{equation}
plus a constant, and the mutual information between $\bm{s}$ and $\bm{x}$ is equal to this entropy up to a constant~\cite{Zhou2023note}. We can take $\mathcal{M} S$ as the total entropy of $\mathcal{M}$ internal states. It is desirable for the entropy $S$ to be large so that the encoding will be sensitive to variations in $\vec{\bm{s}}$ and be robust to noises in representing and transmitting $\vec{\bm{x}}$ (the information maximization principle~\cite{Bell-Sejnowski-1995}).

It turns out that minimizing energy (cost) $E$ and maximizing entropy (information) $S$ are mutually conflicting goals. We introduce a temperature parameter $T$ to quantify this energy--information trade-off. In the thermodynamic limit of infinite number of sensory inputs ($\mathcal{M}\rightarrow \infty$), the optimization objective is the total free energy $\mathcal{M} F$ with the coefficient $F$ being
\begin{equation}
  F = E - T S \; ,
  \label{eq:femin}
\end{equation}
which combines both energetic and entropic effects~\cite{Bell-Sejnowski-1995,Tishby-etal-1999,He-Wang-2017}. The temperature embodies all the external and internal fitness stresses. At high $T$ values information sensitivity and robustness is the main fitness drive and the adaptation of synaptic weights will favor entropy $S$; at the other limit of low $T$, reducing energy $E$ becomes the dominant fitness concern and the weight matrix will evolve towards energy minimization. The thermodynamic equation accompanying free-energy minimization is simply $T = {{\rm d}E}/{{\rm d}S}$.

We search for the global minimum of $F$ at each fixed value of $T$ by a slow simulated annealing process and determine the corresponding optimal weight matrix~\cite{Zhou2023note}. We find that $E(T)$ and $S(T)$ are singular at several critical points of $T$, and the optimal weight matrix $\bm{W}$ changes its qualitative properties at these points. Discontinuities of the energy susceptibility (${\rm d} E/{\rm d}T$) imply continuous phase transitions, and discontinuities in the energy itself manifest discontinuous phase transitions. To demonstrate most explicitly the endogenous nature of such phase transitions, we will focus on symmetric and homogeneous sensory inputs: the self-correlation $c_{i i} = 1$ for all the inputs $s_i$ and pair correlation $c_{i j} = c$ being the same ($c < 1$) for the inputs of any two different units. The default internal model $\bm{W}$ is then reciprocal ($w_{i j} = w_{j i}$) and permutation-symmetric (identical non-diagonal elements)~\cite{Zhou2023note}.

\begin{figure}[t]
  \centering
  \includegraphics[angle=270,width=1.0\linewidth]{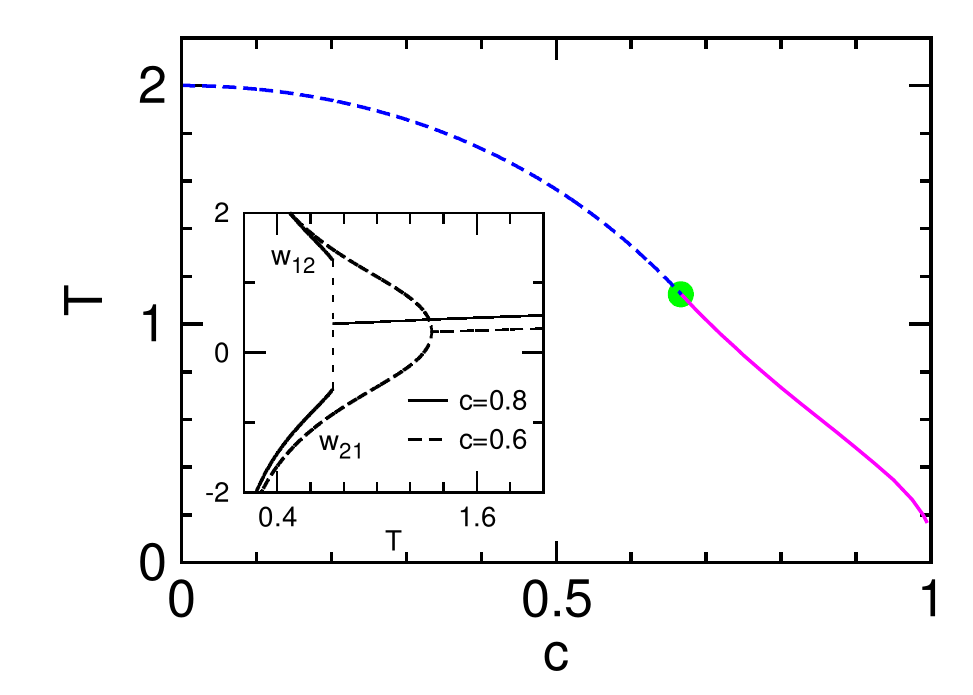}
  \caption{
    Phase diagram for the $N=2$ system. The symmetry-breaking transition to $w_{1 2} \neq w_{2 1}$ is continuous for $c < 2/3$ (blue dashed line) and  discontinuous for $c > 2/3$ (red solid line). The green dot  marks the tricritical point at $c=2/3$, $T= 9/8$. Inset exhibits the continuous transition for $c=0.6$ and the discontinuous one for $c=0.8$.}
  \label{fig:PDN2}
\end{figure}

The phase diagram for the simplest two-unit system (Fig.~\ref{fig:PDN2}) is a good starting point to understand optimal LPC. There is a continuous reciprocity-breaking phase transition for $c < 2/3$, with $w_{1 2}$ and $w_{2 1}$ gradually deviating from each other at the critical temperature $T_2^{\textrm{rb}} = (2-5 c^2/2)/(1-c^2/4)^2$~\cite{Zhou2023note}. When input correlation $c$ exceeds $2/3$, a local minimum of the free energy with $w_{1 2} \neq w_{2 1}$ first emerges in the reciprocal ($w_{1 2}=w_{2 1}$) phase and it then becomes the global minimum as $T$ decreases to certain critical value strictly higher than $T_{2}^{\textrm{rb}}$. A discontinuous transition then occurs, from the reciprocal phase to a reciprocity-broken phase in which unit $2$ inhibits unit $1$ and unit $1$ excites unit $2$.

\subsection{Reciprocity-breaking and ideal-gas law}

\begin{figure}[t]
  \centering
  \subfigure[]{
    \includegraphics[angle=270,width=0.75\linewidth]{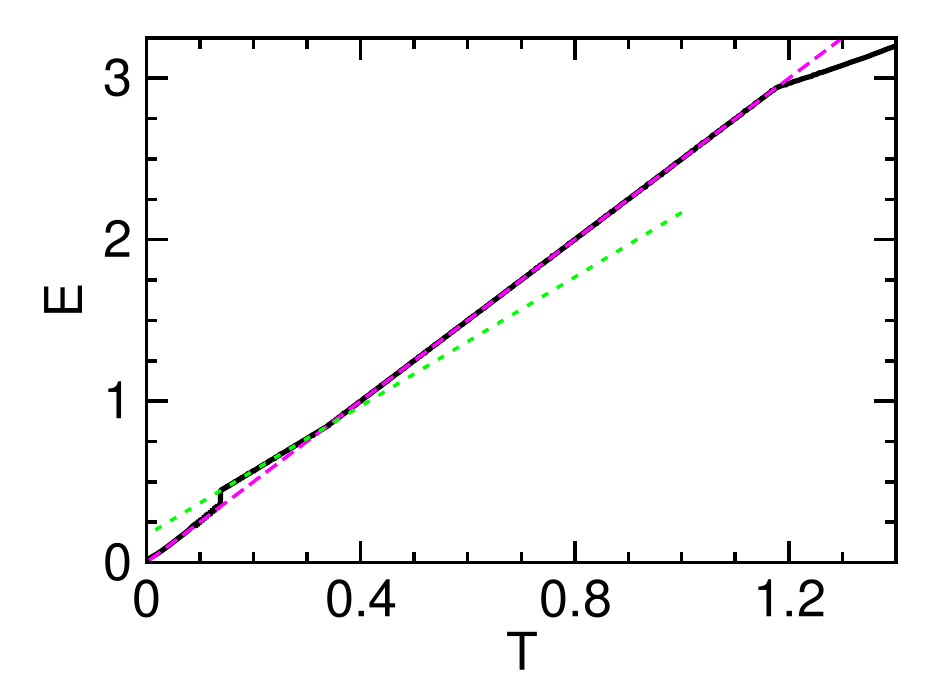}
    \label{fig:N5AWET}
  }
  \subfigure[]{
    \includegraphics[angle=270,width=1.0\linewidth]{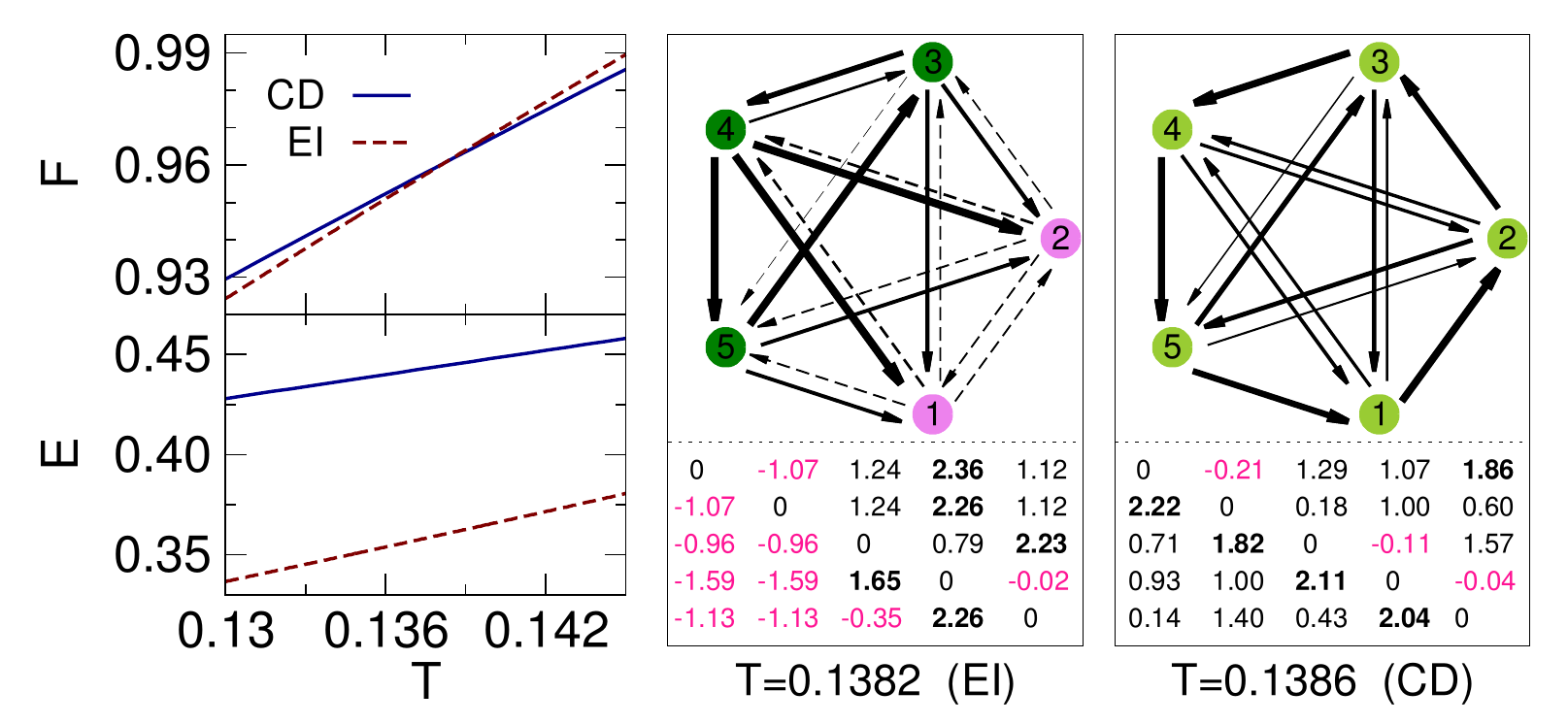}
    \label{fig:N5T138}
  }
  \caption{
    Results for $N=5$ and $c=0.8$. (a) $E$ versus $T$. The thin fitting lines have slopes $2.5$ and $2.0$, respectively (the critical temperatures $T_5^{\textrm{rb}}=1.1783$ and $T_5^{\textrm{iglb}}=0.3360$). (b) Examples of cyclic-dominant and EI-balanced optimal matrices at $T \approx 0.1383$, and $F(T)$ and $E(T)$ for these two branches of solutions. Solid (dashed) links in the interaction graphs indicate positive (negative) weights.}
  \label{fig:N5c0p8AW}
\end{figure}

For multi-unit systems containing $N \geq 3$ units, our numerical optimization results and local stability analysis reveal that the optimal internal model $\bm{W}$ is reciprocal ($w_{i j} = w_{j i}$) and permutation-symmetric (all $w_{i j}$ being equal) at high temperatures, but these properties break down spontaneously  as $T$ drops below the critical value~\cite{Zhou2023note}
\begin{equation}
  T_N^{\textrm{rb}} = 2 \bigl( \sqrt{ 1 + (N - 1) c}
  + (N-1) \sqrt{1 - c} \bigr)^2 /N^2
  \; .
  \label{eq:Tab0}
\end{equation}
However, Eq.~(\ref{eq:Tab0}) does \emph{not} apply for $N=2$, indicating the emergence of new collective properties for $N \geq 3$ at this continuous reciprocity-breaking phase transition. Indeed we find that the energy of the reciprocity-broken ($w_{i j} \neq w_{j i}$) optimal system exactly obeys an ideal-gas law~\cite{Zhou2023note}
\begin{equation}
  E = \frac{N}{2} T   \quad \quad \quad \quad (N \geq 3)
  \label{eq:ideal}
\end{equation}
in an extended temperature interval $(T_N^{\textrm{iglb}}, \,  T_{N}^{\textrm{rb}}]$, with $T_N^{\textrm{iglb}}$ being the lower-bound temperature at which this ideal-gas law is deviated [Fig.~\ref{fig:N5AWET}].

One type of interaction patterns realizing the ideal-gas law is the rotation-symmetric solution with weight parameters $w_i$ (we require $w_1 \leq \ldots \leq w_{N-1}$ to remove trivial degeneracies):
\begin{equation}
  \bm{W} = \begin{bmatrix}
    0  &  w_1  & w_2    & \cdots & w_{N-1} \\
    w_{N-1} & 0      & w_1    & \cdots & w_{N-2} \\
    w_{N-2} & w_{N-1} & 0      & \cdots & w_{N-3} \\
    \vdots & \vdots & \vdots & \ddots & \vdots \\
    w_1    & w_2    & w_3    & \cdots & 0
  \end{bmatrix}
  \; .
\end{equation}
There is cyclic dominance (CD) among the $N$ units such that unit $i+1$ is most strongly inhibited by unit $i$ and it most strongly inhibits unit $i+2$ and so on. Rotation-symmetric solutions are impossible for $N=2$, and there is only one such solution for $N=3, 4$ but infinitely many degenerate solutions for $N\geq 5$~\cite{Zhou2023note}.

We can define a quantity $\mathcal{O}_{CD}$, which compares the most dominant global cycle of directed interactions and the reverse directed cycle of interactions, to measure the degree of cyclic dominance:
  \begin{equation}
    \mathcal{O}_{CD} = 1 - \min\limits_{(p_1,\ p_2,\ p_N)} \biggl|
    \frac{ \sum_{i=1}^{N-1} w_{p_{i+1} \ p_i}  }{
      \sum_{i = 1}^{N-1} w_{p_i\ p_{i+1} } } \biggr|
    \; ,
  \end{equation}
  where $(p_1, p_2, \ldots, p_N)$ denotes a permutation of the $N$ units. For example, the weight matrix shown in the right panel of Fig.~\ref{fig:N5T138} has a very high value of $\mathcal{O}_{CD}=0.996$. We find through numerical computations that the ideal-gas law can be achieved by cyclic-dominant weight matrices with or without rotational symmetry and also by two-component matrices composed of excitatory and inhibitory units~\cite{Zhou2023note}.

  What is the significance of the ideal-gas law (\ref{eq:ideal})? In terms of the eigenvalues $\epsilon_i$ of the symmetric matrix $(\bm{I}+\bm{W})^{-1} C (\bm{I}+\bm{W}^\top)^{-1}$ we have $E = \sum_i \epsilon_i$ and $2 S= \sum_i \ln \epsilon_i - \ln \textrm{Det}(\bm{C})$. The entropy is then upper-bounded by $S \leq (N/2) \ln (E/N) - (1/2) \ln \textrm{Det}(\bm{C})$ at each energy $E$, and equality holds only if all the eigenvalues $\epsilon_i$ are identical~\cite{Zhou2023note}. If this entropy upper-bound can be saturated within a continuous interval of $E$, then Eq.~(\ref{eq:ideal}) will be achieved as a consequence of energy--information competition, and it means the most efficient utilization of energy to reach the information robustness upper-bound. An appealing statistical property is that the internal states are composed of independent components of equal magnitudes~\cite{Jutten-Herault-1991,Hyvarinen-Oja-2000}, with $\langle x_i^2 \rangle =  E/N$ to achieve energy equipartition and $\langle x_i x_j \rangle = 0$ for $j \neq i$ to eliminate internal pair correlations (here $\langle \cdot \rangle$ denotes averaging over all the samples). We notice a probably closely related biological fact is that the neuron pair correlations of the visual cortex are very weak even for highly correlated sensory inputs~\cite{Schneidman-etal-2006}.

As the temperature decreases to the lower critical value $T_N^{\textrm{iglb}}$, it will no longer be possible for an optimal weight matrix to exactly obey Eq.~(\ref{eq:ideal}) and at the same time keep all its complex eigenvalues above the bottomline. Then the ideal-gas law  will break down and a kink of the energy function $E(T)$ will be observed.  For $N=5$ and $c=0.8$, $T_5^{\textrm{iglb}}=0.3360$ at which the energy slope changes from $2.5$ to $2.0$ [Fig.~\ref{fig:N5AWET}]. This continuous transition is not associated with any symmetry change.

\subsection{Discontinuous phase transitions}

The energy discontinuity at $T=0.1383$ in the example curve of Fig.~\ref{fig:N5AWET} signifies a discontinuous phase transition. Indeed the free energy at the vicinity of this temperature has two minima and they organize into two branches with distinct energies [Fig.~\ref{fig:N5T138}]. One branch corresponds to the one-component CD network which is almost fully inhibitory. The other branch corresponds to a two-component EI-balanced network: there is permutation symmetry within the two excitatory units $1$ and $2$ of group $g_E$, while the three units $3,4,5$ of group $g_I$ have cyclically dominant interactions within themselves and they strongly inhibit group $g_E$. Such a pattern is in agreement with the vital biological fact of excitation--inhibition competition and balance~\cite{vanVreeswijk-Sompolinsky-1996,Yang-Zhou-etal-2017,Yu-etal-2018}.

One simple measure of EI-balance is to compute the net input weights of all the individual units,
  \begin{equation}
    \mathcal{O}_{EI}
    = \frac{1}{N} \sum\limits_{i = 1}^{N}
    \biggl[ 1 -  \frac{ \bigl| \sum_{j \neq i} w_{i j} \bigr|}{
        \sum_{j \neq i} |w_{i j} |} \biggr] \; .
  \end{equation}
  A value of $\mathcal{O}_{EI}$ significantly above zero indicates a highly EI-balanced network in whcih the excitatory inputs are largely cancelled by the inhibitory inputs for most of the units.  For example,  $\mathcal{O}_{EI}=0.626$ for the weight matrix shown in the middle panel of Fig.~\ref{fig:N5T138}.

\begin{figure}[t]
  \centering
  \includegraphics[angle=270,width=1.0\linewidth]{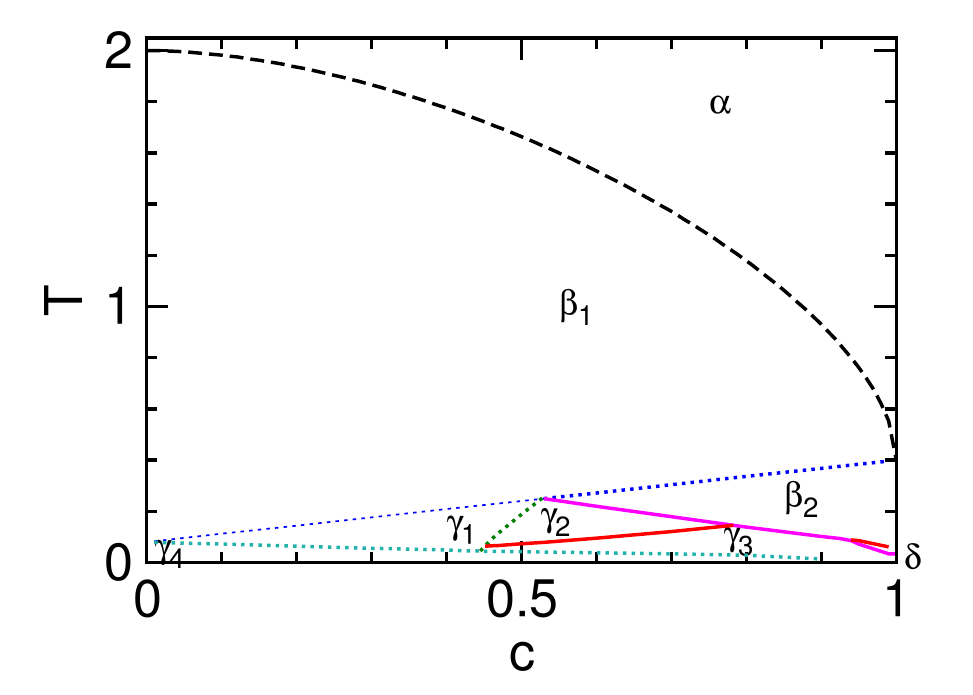}
  \caption{
    Phase diagram for $N=5$. The dashed line obeys Eq.~(\ref{eq:Tab0}); solid lines indicate discontinuous transitions, and dotted lines mark continuous transitions attributed to the eigenvalue bottomline constraint. Region $\alpha$: permutation-symmetric; $\beta_1$--$\beta_2$: cyclic dominant; $\gamma_1$--$\gamma_4$ (and part of $\beta_1$): EI-balanced; $\delta$: an intermediate phase. $\beta_1$ and $\gamma_1$ obey the ideal-gas law Eq.~(\ref{eq:ideal}).
  }
  \label{fig:PDN5AW}
\end{figure}

After the transition to the optimal EI-balanced network, the ideal-gas law is then approximately (though not exactly) recovered at lower temperatures [Fig.~\ref{fig:N5AWET}], indicating that EI-balance is beneficial for the efficiency of energy usage. This can not be achieved if all the synaptic weights are restricted to be positive (inhibitory)~\cite{Zhou2023note}.

The phase diagram for the system containing $N=5$ units is shown in Fig.~\ref{fig:PDN5AW}. The $\beta_1$ and $\beta_2$ regions in this diagram correspond to one-component cyclic-dominant optimal networks (a subregion of $\beta_1$ also contains EI-balanced ones). The $\gamma_1$--$\gamma_4$ regions all correspond to different types of EI-balanced optimal networks. The phase boundaries are precisely determined by the singularities of $E(T)$, and we can also define some quantitative measures such as $\mathcal{O}_{CD}$ and $\mathcal{O}_{EI}$ to describe these different phases~\cite{Zhou2023note}. The ideal-gas law is exact in regions $\beta_1$ and $\gamma_1$ but it only approximately holds in $\gamma_2$--$\gamma_4$. The discontinuous transition from $\gamma_2$ to $\gamma_3$ is associated with breaking the internal rotational symmetry of group $g_I$ and it leads to functional differentiation of the three units $3, 4, 5$ (see Ref.~\cite{Zhou2023note} for more details). The phase diagram for the smallest multi-unit system ($N=3$) is qualitatively similar to Fig.~\ref{fig:PDN5AW}; and we also report some numerical results on the larger system of $N=10$ units to demonstrate the generality of our conclusions~\cite{Zhou2023note}.

\subsection{Discussion}

Our theoretical work attempted to consider energy and function simultaneously in a single model, and revealed that the trade-off between energy reduction and information robustness can induce spontaneous breaking of permutation symmetry in the lateral predictive coding system and drive the formation of cyclic dominance and the sudden emergence of excitation--inhibition balance among the network units. We discovered an ideal-gas law (\ref{eq:ideal}) of internally representing the sensory inputs as equal-magnitude independent components. Our theoretical results are in qualitative agreement with some important biological facts such as non-reciprocal interactions, excitation--inhibition balance and weak internal pair correlations. They may also have algorithmic implications.

For simplicity we have assumed the input correlation matrix $\bm{C}$ is uniform with a single parameter $c$. A natural extension of the present work is to study the ideal-gas law (\ref{eq:ideal}) for heterogeneous sensory inputs. Our ongoing analytical and numerical computations confirm that Eq.~(\ref{eq:ideal}) and continuous and discontinuous phase transition will also be observed for non-uniform matrices $\bm{C}$~\cite{Zhou2023note,Huang-etal-2023}.

The existence of discontinuous phase transitions and the associated multiple free energy minima may render the optimal weight matrices hard to acquire through local and gradual Hebbian learning processes~\cite{Fumarola-etal-2022,Pineda-1987,Spratling-2017}. This issue waits to be further explored.  It is also very interesting to study phase transitions in nonlinear LPC models with energy cost being a sum $\sum_i | x_i |$ of absolute values and internal prediction being rectified linear ${\rm max}\bigl(0, \sum_{i\neq j} w_{i j} x_j\bigr)$ or another more complicated nonlinear function~\cite{Curto-etal-2019}. We expect to observe rich continuous and discontinuous phase transitions in these more realistic models too. These nonlinear LPC systems may lead to enhanced sensitivity to fine details in typical input vectors.  

\begin{acknowledgments}
  The following funding supports are acknowledged: National Natural Science Foundation of China Grants No. 11421063, No. 11747601 and No.~12247104; National Innovation Institute of Defense Technology Grant No.~22TQ0904ZT01025. Numerical simulations were carried out at the HPC cluster of ITP-CAS and also at the BSCC-A3 platform of the National Supercomputer Center in Beijing with the help of the TRNG random number generators~\cite{Bauke-Mertens-2007}.
\end{acknowledgments}

\clearpage

\begin{widetext}

  \begin{center}
    \textbf{Supplementary Information}
  \end{center}
  \vskip 1.0cm
  
    Supplementary Information for the main text. It contains some detailed theoretical derivations, brief descriptions of the numerical simulation methods, and some additional numerical results.
 
  The equations in this supplementary text are indexed following those of the main text. Same for the figures.

\subsection{On the definition of entropy}

Suppose there is noise in the mapping from sensory input $\bm{s} = (s_1, \ldots, s_N)^\top$ to the internal representation $\bm{x} = (x_1, \ldots, x_N)^\top$,
\begin{equation}
  x_i = g_i(\bm{s}) + \xi_i \; ,
\end{equation}
where $\xi_i$ for $i=1, \ldots, N$ are independent Gaussian random variables with mean zero and variance $\sigma^2$. For the linear lateral predictive coding (LPC) system, the deterministic function $g_i(\bm{s})$ is
\begin{equation}
  g_i( \bm{s} ) = \sum_{j=1}^N \Bigl( \frac{\bm{I}}{\bm{I} + \bm{W}} \Bigr)_{i j} s_j \; ,
\end{equation}
where $\bm{I}$ is the identity matrix with $I_{i i} = 1$ and $I_{i j} = 0$ for any $i \neq j$. The mutual information between $\bm{x}$ and $\bm{s}$ is
\begin{equation}
  \mathcal{I}[\,\bm{x}; \bm{s}\,] = H[ \bm{x} ] + H[ \bm{s} ] -
  H[\, \bm{x} | \bm{s}\, ]
  \; ,
\end{equation}
where $H[\bm{x}]$ and $H[\bm{s}]$ are the entropies of internal states $\bm{x}$ and sensory inputs $\bm{s}$, respectively, and $H[\, \bm{x} | \bm{s}\, ]$ is the conditional entropy of $\bm{x}$ given $\bm{s}$. Notice that the entropy $H[ \bm{s} ]$ only depends on the probability distribution $P_{\textrm{input}}(\bm{s})$ of the sensory inputs $\bm{s}$, so it could be considered as a constant. The conditional entropy $H[\,\bm{x} | \bm{s}\,]$ only depends on the property of the noise vector $(\xi_1, \ldots, \xi_N)^\top$ and so it could also be regarded as a constant. Then the mutual information $\mathcal{I}[\, \bm{x} ; \bm{s}\, ] = H[ \bm{x} ] + \mathcal{I}_0$ with $\mathcal{I}_0$ being a constant. To increase this mutual information we need to increase the entropy $H[\bm{x}]$ of internal states $\bm{x}$.

The marginal distribution $P_{\textrm{output}}(\bm{x})$ of the internal state $\bm{x}$ is
\begin{eqnarray}
  P_{\textrm{output}}( \bm{x} ) &  = & \int
  {\rm d} \bm{s} P_{\textrm{input}}( \bm{s} ) 
  \prod\limits_{i=1}^N
  \Bigl[ \frac{1}{\sqrt{ 2 \pi \sigma^2}} \exp\bigl(
    - \frac{(x_i - g_i(\bm{s}) )^2}{2 \sigma^2}
    \bigr) \Bigr]
  \\
  & \approx & \Bigl| \textrm{Det} \bigl( \bm{I} + \bm{W} \bigr) \Bigr| \;
  P_{\textrm{input}}\bigl( (\bm{I} + \bm{W} ) \bm{x} \bigr) \; .
  \label{eq:pexpress}
\end{eqnarray}
Equation (\ref{eq:pexpress}) is valid for the situation of small noises ($\sigma^2$ being sufficiently small). The entropy of $\bm{x}$ is then
\begin{equation}
  H[\bm{x}] \equiv  - \int {\rm d} \bm{x} P_{\textrm{output}}( \bm{x})
  \ln P_{\textrm{output}}(\bm{x} ) = - \ln \Bigl| \textrm{Det} \bigl( \bm{I} + \bm{W} \bigr) \Bigr| + H[ \bm{s} ]
  \; .
\end{equation}
Again because $H[\bm{s}]$ does not depend on the weight matrix $\bm{W}$, we see that $H[\bm{x}] = -\ln\bigl| \textrm{Det}(\bm{I}+\bm{W}) \bigr| + H_0$ with $H_0$ being a constant. For our LPC problem, the real parts of all the eigenvalues of $\bm{I}+\bm{W}$ are required to be positive. The determinant of $\bm{I}+\bm{W}$ then must be positive. This explains Eq.~(3) of the main text.

\subsection{The two-unit system ($N=2$)}

When there are only two units, the steady-state relationship between the sensory input $\bm{s}$ and the internal state $\bm{x}$ is
\begin{equation}
  x_1  = \frac{s_{1} - w_{1 2}\, s_2}{1 - w_{1 2}\, w_{2 1}} \; ,
  \quad
  x_2  = \frac{s_2 - w_{2 1} \, s_1}{1 - w_{1 2}\, w_{2 1}} \; .
\end{equation}
The mean energy of the system is
\begin{equation}
  E  =  \bigl\langle (x_1)^2 \bigr\rangle + \bigl\langle (x_2)^2 \bigr\rangle
  =  \frac{c_{1 1} + c_{2 2} - 2 c_{1 2}\, (w_{1 2} + w_{2 1})
    + c_{2 2}\, w_{1 2}^2 + c_{1 1}\, w_{2 1}^2}{(1- w_{1 2}\, w_{2 1})^2}
  \; .
\end{equation}
Here $\langle \cdot \rangle$ denotes averaging over all the input samples, and $c_{i j}$ is the element of the input correlation matrix $\bm{C}$: 
\begin{equation}
  c_{1 1} = \langle s_1^2 \rangle \; ,
  \quad 
  c_{2 2} = \langle s_2^2 \rangle \; ,
  \quad 
  c_{1 2} = \langle s_1 s_2 \rangle \; .
\end{equation}
The entropy of the system is
\begin{equation}
  S = - \ln \bigl( 1 - w_{1 2}\, w_{1 2} \bigr) \; ,
\end{equation}
which only depends on the product of the two synaptic weights.

We can determine the minimum value of energy $E$ at each fixed value of $w_{1 2}\, w_{2 1}$. Some example curves of $E$ versus $w_{1 2}\, w_{2 1}$ are shown in Fig.~\ref{fig:221024a} for statistically symmetric inputs with $c_{1 1} = c_{1 2} = 1$, and the corresponding synaptic weights $w_{1 2}$ and $w_{2 1}$ are shown in Fig.~\ref{fig:221022a}. We see that $w_{1 2} \neq w_{2 1}$ when the value $w_{1 2} \, w_{2 1}$ is below certain threshold value, at each fixed value $c_{1 2}$ of the input correlation. 

\begin{figure}[h]
  \centering
  \subfigure[]{
    \includegraphics[angle=270,width=0.45\textwidth]{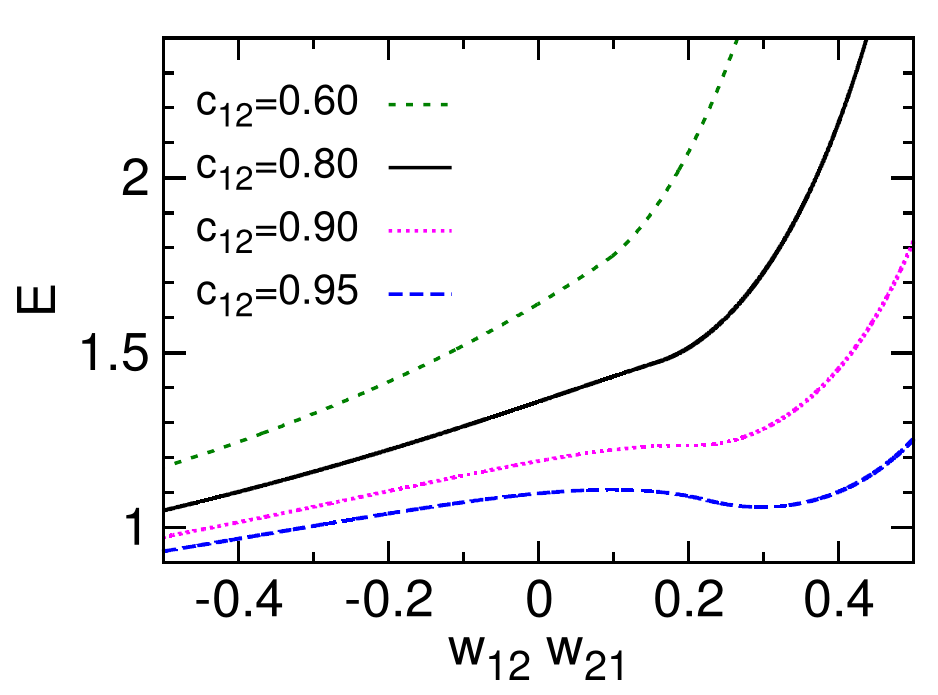}
    \label{fig:221024a}
  }
  \subfigure[]{
    \includegraphics[angle=270,width=0.45\textwidth]{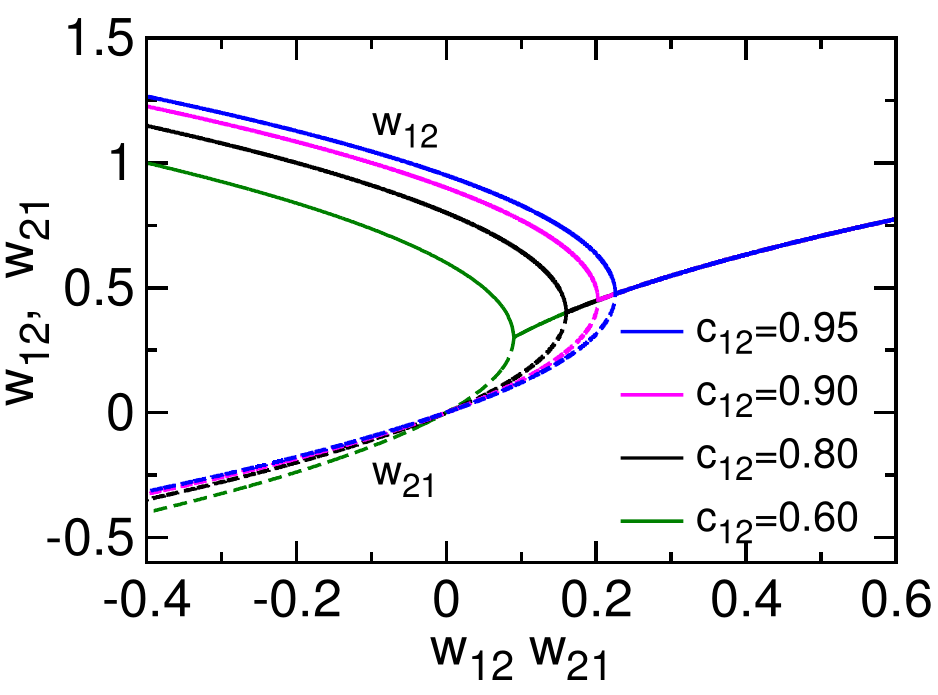}
    \label{fig:221022a}
  }
  \caption{
    Minimum energy value $E$ (a) and the weights $w_{1 2}$ and $w_{2 1}$ (b) as a function of the product $w_{1 2}\, w_{2 1}$ for the symmetric $N=2$ system with $c_{1 1} = c_{2 2}=1$ and $c_{1 2} = 0.60,\; 0.8,\; 0.9,\; 0.95$.
  }
  \label{fig:weightN2}
\end{figure}

The minimum value of the free energy $F$ is determined at each value of the temperature $T$ by choosing the optimal value of $w_{1 2}\, w_{2 1}$. Then the energy $E$ as a function of $T$ can be obtained. Some example results are shown in Fig.~\ref{fig:N2EW} for symmetric inputs ($c_{1 1}  = c_{2 2} = 1$).

\begin{figure}[h]
    \centering
    \subfigure[]{
      \includegraphics[angle=270,width=0.465\linewidth]{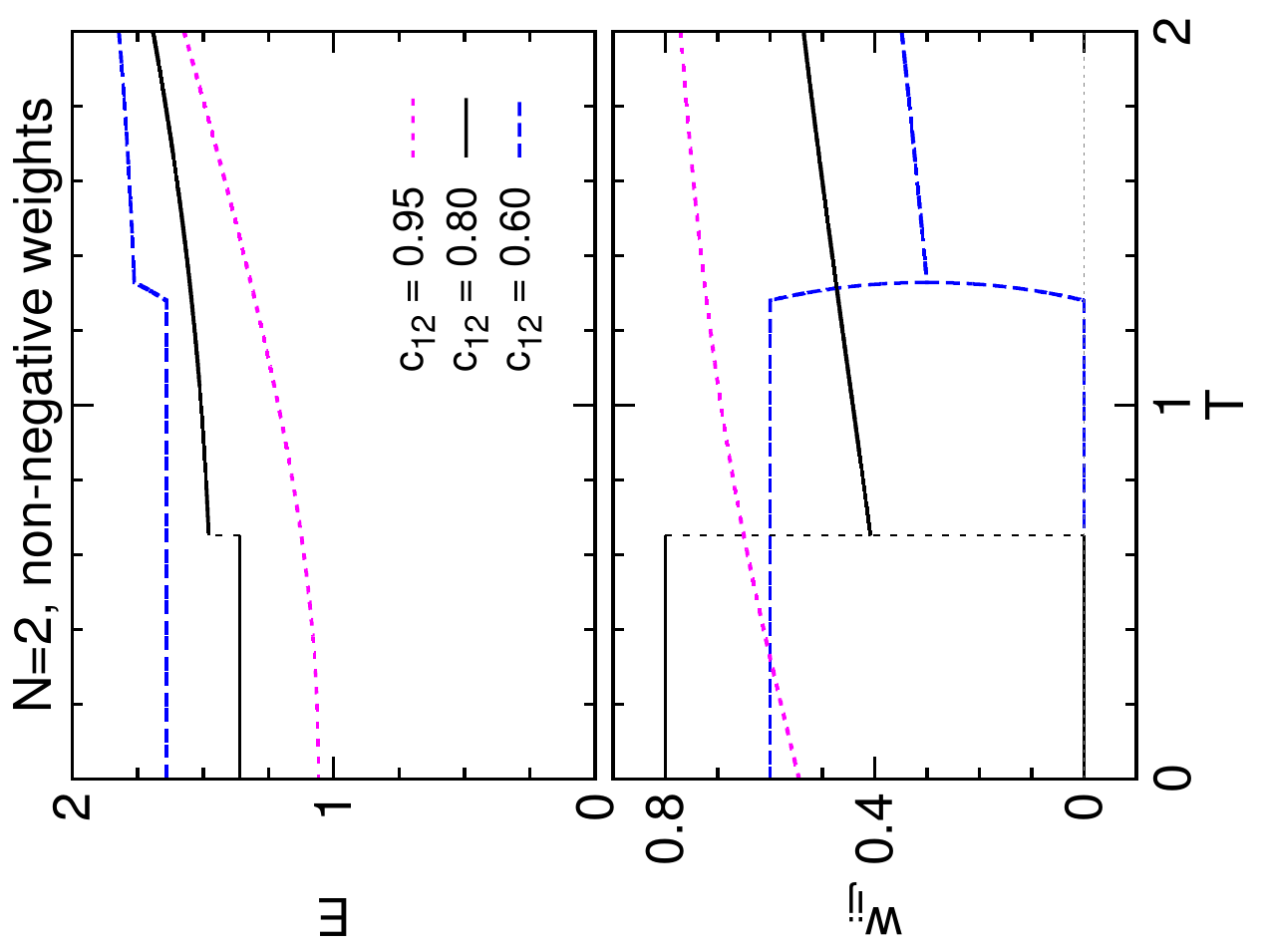}
    \label{fig:N2EW:nnw}
    }
    \subfigure[]{
      \includegraphics[angle=270,width=0.465\linewidth]{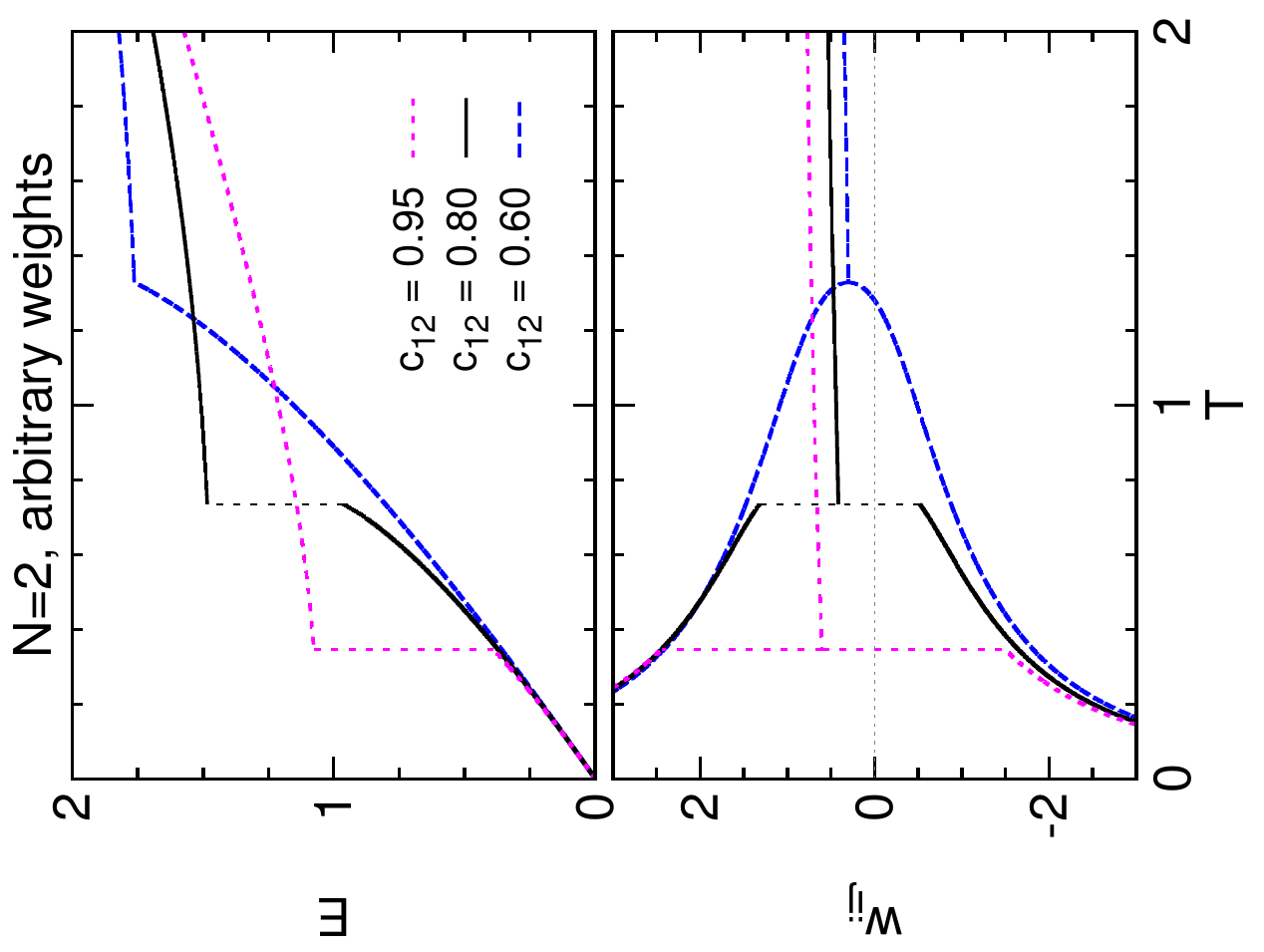}
      \label{fig:N2EW:aw}
    }
    \caption{
      Optimal predictive coding for the two-unit system with symmetric inputs ($c_{1 1} = c_{2 2} = 1$). Synaptic weights are restricted to be non-negative in (a) but are unrestricted in (b). Mean energy $E$ and weights $w_{1 2}$ and $w_{2 1}$ are shown for three representative values of input pair correlation $c_{1 2} = 0.60$ (blue long dashed line), $0.80$ (black solid line), and $0.95$ (red dashed line).
    }
    \label{fig:N2EW}
\end{figure}

If the synaptic weights are restricted to be inhibitory ($w_{i j} \geq 0$), then depending on the input pair correlation $c_{1 2}$, there are three different outcomes [Fig.~\ref{fig:N2EW:nnw}]. If $c_{1 2}$ is close to unity (say $0.95$), synaptic reciprocity $w_{1 2}=w_{2 1}$ is always kept and both $E$ and $S$ are smooth functions of $T$. At lower values of $c_{1 2}$ (e.g., $0.80$), interaction reciprocity breaks down discontinuously at certain critical value of $T$, with an associated sudden drop in the energy $E$ and in the entropy $S$.  At still lower values of $c$ (e.g., $0.60$), the breaking of reciprocity becomes a continuous transition, and both $E$ and $S$ are continuous at the transition point but their first-order derivatives are discontinuous. The minimum-energy network will be reached when one of the synaptic weights vanishes. The minimum energy can not be arbitrarily small, because the input to one unit (say $s_1$) is not cancelled by the output $x_2$ of the other unit.

If the synaptic weights are allowed to take negative (excitatory) values, the mean energy of the network at low temperature values can be remarkably reduced [Fig.~\ref{fig:N2EW:aw}].  The optimal strategy to achieve sufficiently low energy $E$ appears to be that one unit inhibits and is excited by the other unit. There is either a continuous or a discontinuous phase transition for $c \leq \frac{2}{3}$ and $c > \frac{2}{3}$, respectively. The phase diagram for this case of arbitrary weights is reported in Figure 1 of the main text. 

When $c_{1 1} \neq c_{2 2}$ there is no permutation symmetry to break and the continuous phase transition is absent. We find that the discontinuous phase transition is still possible for this non-symmetric case. Setting $c_{2 2} = 1$, Fig.~\ref{fig:N2A12A11} shows the region of the $c_{1 1}$--$c_{1 2}$ plane in which the discontinuous phase transition can be observed at certain critical temperature.

\begin{figure}[h]
  \centering
  \includegraphics[width = 0.5\linewidth]{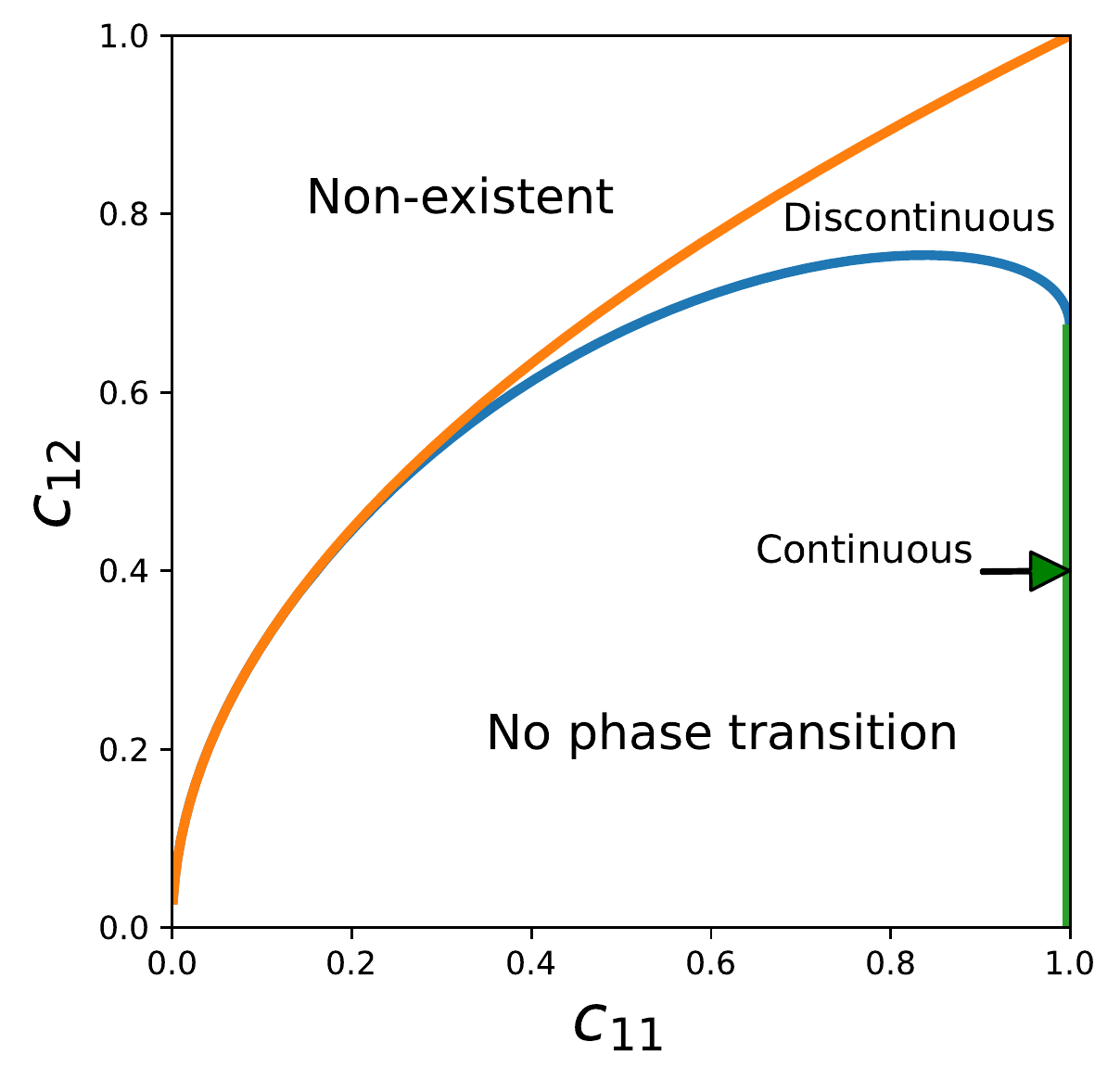}
  \caption{
    Discontinuous phase transition is possible for the $N=2$ system with $c_{2 2} = 1$ and $c_{1 1} < 1$ in the region between the two lines of this $c_{1 1}$--$c_{1 2}$ plane. The upper-boundary line corresponds to the limiting value $c_{1 2} = \sqrt{c_{1 1}}$. The lower-boundary line is determined numerically.
  }
  \label{fig:N2A12A11}
\end{figure}

\subsection{Local stability of the reciprocal and permutation-symmetric solution}

For symmetric input vectors $\bm{s}$ with $c_{i i} \equiv \langle s_i^2 \rangle = 1$ and $c_{i j} \equiv \langle s_i s_j \rangle = c$ for any $i \neq j$, the corresponding symmetric internal model is the permutation-symmetric matrix $\bm{W}$ with all its non-diagonal elements being equal to $w$. Reciprocity obviously holds in this internal model ($w_{i j} = w_{j i}$ for all pairs). We now determine the value of this uniform weight $w$ and study the local stability of this reciprocal and permutation-symmetric solution.

First, notice that the homogeneous correlation matrix $\bm{C}$ could be written as
\begin{equation}
  \bm{C} =  (1 - c)\, \bm{I} + c\,
  \begin{bmatrix}
    1 \\
    1\\
    \vdots \\
    1
  \end{bmatrix}
  \,
  \begin{bmatrix}
    1 & \;1  & \; \ldots & \; 1
  \end{bmatrix}
  \; ,
\end{equation}
from which it is easy to see that $\bm{C}$ has one eigenvalue $1+(N-1) c$ with the corresponding eigenvector $\bm{u}_0=\frac{1}{\sqrt{N}}\bigl(1, 1, \ldots, 1\bigr)^\top$ and $N-1$ degenerate eigenvalues $(1-c)$ and the corresponding $N-1$ eigenvectors which are orthogonal to $\bm{u}_0$ and to each other. The symmetric matrix $\bm{I} + \bm{W}$ can be diagonalized in the same way. Since all the eigenvalues of the correlation matrix must be non-negative, so $c \in [- \frac{1}{N-1}, \; 1 ]$. To ensure that the short-time predictive coding dynamics be convergent, the eigenvalues of $\bm{I}+\bm{W}$ are required to be positive, which means $w \in ( - \frac{1}{N-1}, \; 1 )$. When the matrix $\bm{W}$ is permutation-symmetric, the entropy $S$ is expressed as
\begin{equation}
  S =  - \ln\bigl( 1+ (N-1) w \bigr) - (N-1) \ln(1 - w ) \; .
\end{equation}

Let us define an auxiliary symmetric matrix $\bm{M}$ for a general weight matrix $\bm{W}$ as
\begin{equation}
  \bm{M} = \frac{\bm{I}}{\bm{I}+\bm{W}} \bm{C}
  \Bigl( \frac{\bm{I}}{\bm{I}+\bm{W}}\Bigr)^\top \; .
  \label{eq:matrixM}
\end{equation}
In the special case of $\bm{W}$ being permutation-symmetric, it is easy to check that
\begin{equation}
  \bm{M}   = \frac{1-c}{(1-w)^2}\, \bm{I}
  + \frac{1}{N} \Bigl( \frac{1+(N-1)c}{(1+(N-1) w)^2}
  - \frac{1-c}{(1-w)^2} \Bigr)\,
  \begin{bmatrix}
    1\\
    1\\
    \vdots \\
    1
  \end{bmatrix}
  \, 
  \begin{bmatrix}
    1 \; & 1\; & \ldots\;  & 1
  \end{bmatrix}
  \; ,
\end{equation}
and then we obtain the following expression for the energy:
\begin{equation}
  E  \equiv   \textrm{Tr} \bigl[ \bm{M} \bigr] 
  = \frac{1+(N-1) c}{ \bigl( 1+(N-1) w \bigr)^2} 
  +(N-1) \frac{1-c}{(1-w)^2 } \; .
\end{equation}
By minimizing the free energy $F = E - T S$ with respect to the weight parameter $w$, we obtain the following relationship between $w$ and the temperature $T$:
\begin{equation}
  T  = \frac{2}{N w} \Bigl( \frac{(1-c) \bigl(1 + (N-1) w \bigr)}{(1-w)^2} -
  \frac{(1-w) \bigl(1 + (N-1) c\bigr)}{[ 1 + (N-1) w]^2}
  \Bigr) \; .
  \label{eq:Tpsfix}
\end{equation}

The local stability of this symmetric solution is studied by assuming $w_{i j} = w + \varepsilon_{i j}$ with $\varepsilon_{i j}$ being a tiny perturbation of weight $w_{i j}$. Denote by $\bm{\varepsilon}$ the perturbation matrix whose diagonal elements are all being exactly zero, we have
\begin{equation}
  \frac{\bm{I}}{\bm{I}+\bm{W} + \bm{\varepsilon}} =
  \frac{\bm{I}}{\bm{I}+\bm{W}}
  - \frac{\bm{I}}{\bm{I} + \bm{W}} \bm{\varepsilon}
  \frac{\bm{I}}{\bm{I} + \bm{W}}
  + \frac{\bm{I}}{\bm{I} + \bm{W}} \bm{\varepsilon}
  \frac{\bm{I}}{\bm{I} + \bm{W}} \bm{\varepsilon}
  \frac{\bm{I}}{\bm{I} + \bm{W}} + \ldots \; .
\end{equation}
Then the energy deviation $\Delta E$ up to the second order of $\bm{\varepsilon}$ is
\begin{eqnarray}
  \Delta E & = &
  - \textrm{Tr}\Bigl[
    \bm{M} \bm{\varepsilon}^\top \frac{\bm{I}}{\bm{I}+\bm{W}^\top}
    +  \frac{\bm{I}}{\bm{I}+\bm{W}}
    \bm{\varepsilon} \bm{M} \Bigr]
  \nonumber \\
  & & + \textrm{Tr}\Bigl[ \bm{M} \bm{\varepsilon}^\top
    \frac{\bm{I}}{\bm{I}+\bm{W}^\top}
    \bm{\varepsilon}^\top \frac{\bm{I}}{\bm{I}+\bm{W}^\top}
     + \frac{\bm{I}}{\bm{I}+\bm{W}}
    \bm{\varepsilon} \frac{\bm{I}}{\bm{I}+\bm{W}}
    \bm{\varepsilon} \bm{M}
    \nonumber \\
    & & \quad \quad \quad 
    +  \frac{\bm{I}}{\bm{I}+\bm{W}}
    \bm{\varepsilon} \bm{M}
    \bm{\varepsilon}^\top \frac{\bm{I}}{\bm{I}+\bm{W}^\top}
    \Bigr] \; .
\end{eqnarray}
The first and second derivatives of the entropy are, respectively,
\begin{eqnarray}
  \frac{ \partial S}{\partial w_{i j} } & = & 
  - \Bigl( \frac{\bm{I}}{\bm{I}+\bm{W} } \Bigr)_{j i} \; , \\
  \frac{ \partial^2 S}{\partial w_{i j} \partial w_{k l} }
  & = & - \frac{\partial}{\partial w_{k l}}
  \Bigl( \frac{\bm{I}}{\bm{I}+\bm{W} } \Bigr)_{j i}
  \; = \; 
  \Bigl( \frac{\bm{I}}{\bm{I}+\bm{W} } \Bigr)_{j k} \; \Bigl( \frac{\bm{I}}{\bm{I}+\bm{W} } \Bigr)_{l i} \; ,
\end{eqnarray}
and therefore the deviation of entropy up to second order is
\begin{equation}
  \Delta S = 
  - \textrm{Tr} \Bigl[ \bm{\varepsilon} \frac{\bm{I}}{\bm{I} + \bm{W}} \Bigr]
  + \frac{1}{2} \textrm{Tr}\Bigl[ \bm{\varepsilon}
    \frac{\bm{I}}{\bm{I}+\bm{W}} \bm{\varepsilon}
    \frac{\bm{I}}{\bm{I} + \bm{W}} \Bigr] \; .
\end{equation}

The first-order deviation ($\Delta F^{(1)}$) of the free energy is then
\begin{equation}
  \Delta F^{(1)} = 
  -  \textrm{Tr}\Bigl[ \bm{M} \bm{\varepsilon}^\top
    \frac{\bm{I}}{\bm{I}+\bm{W}^\top} +  \frac{\bm{I}}{\bm{I}+\bm{W}}
    \bm{\varepsilon} \bm{M} \Bigr]
  + T\, \textrm{Tr} \Bigl[ \bm{\varepsilon} \frac{\bm{I}}{\bm{I} + \bm{W}}
    \Bigr]
  \; .
  \label{eq:fefirst}
\end{equation}
If this first-order deviation is identical to zero for all perturbation matrices $\bm{\varepsilon}$ which do not violate the eigenvalue constraints of $\bm{W}$, then the matrix $\bm{W}$ must be an extremal point of the free energy $F$. The permutation-symmetric matrix with $w$ satisfying Eq.~(\ref{eq:Tpsfix}) is such an extremal solution.

The second-order derivation ($\Delta F^{(2)}$) of the free energy is expressed as
\begin{equation}
  \Delta F^{(2)} = \sum\limits_{i, j} \sum\limits_{k, l} L_{i j, k l} \varepsilon_{i j} \varepsilon_{k l} \; ,
\end{equation}
where  $i \neq j$ and $k \neq l$ and the element $L_{i j, k l}$ of the symmetric Hessian matrix $\bm{L}$ is computed according to
\begin{eqnarray}
  L_{i j, k l} & = & \sum\limits_{m} \Bigl[
    \Bigl( \frac{\bm{I}}{\bm{I}+\bm{W}} \Bigr)_{l i} \Bigl( \frac{\bm{I}}{\bm{I}+\bm{W}} \Bigr)_{m k} M_{m j} 
    + \Bigl( \frac{\bm{I}}{\bm{I}+\bm{W}} \Bigr)_{j k} \Bigl( \frac{\bm{I}}{\bm{I}+\bm{W}} \Bigr)_{m i} M_{m l}
  \nonumber \\
  & & \quad \quad 
  +  \Bigl( \frac{\bm{I}}{\bm{I}+\bm{W}} \Bigr)_{m i} \Bigl( \frac{\bm{I}}{\bm{I}+\bm{W}} \Bigr)_{m k} M_{j l} \Bigr]
  - \frac{T}{2}  \Bigl( \frac{\bm{I}}{\bm{I}+\bm{W}} \Bigr)_{j k} \Bigl( \frac{\bm{I}}{\bm{I}+\bm{W}} \Bigr)_{l i}
  \; .
  \label{eq:Lhessian}
\end{eqnarray}

When the minimum eigenvalue of this $N(N-1) \times N (N-1)$ symmetric matrix $\bm{L}$ changes from being positive to being negative, the synaptic matrix $\bm{W}$ will no longer be a locally stable solution of the free energy minimization problem. For the permutation-symmetric matrix $\bm{W}$ the Hessian matrix is easy to compute through Eq.~(\ref{eq:Lhessian}). Our numerical results for the cases of $N=2, 3, 4, 5, 10$ are shown in Fig.~\ref{fig:localstab}. This reciprocity-breaking (and also permutation-symmetry-breaking) temperature is denoted as $T_{N}^{\textrm{rb}}$ in the main text.

When $N=2$ the Hessian matrix $\bm{L}$ is a $2\times 2$ matrix, and it is straightforward to obtain the analytical expression of $T_{2}^{\textrm{rb}}$ as
\begin{equation}
  T_{2}^{\textrm{rb}} = \frac{2 - 5 c^2 / 2}{ (1 - c^2/4)^2 }
  \; ,
\end{equation}
which is in full agreement with the numerical results obtained at $N=2$.

\begin{figure}[h]
\centering
\includegraphics[angle=270,width=0.5\textwidth]{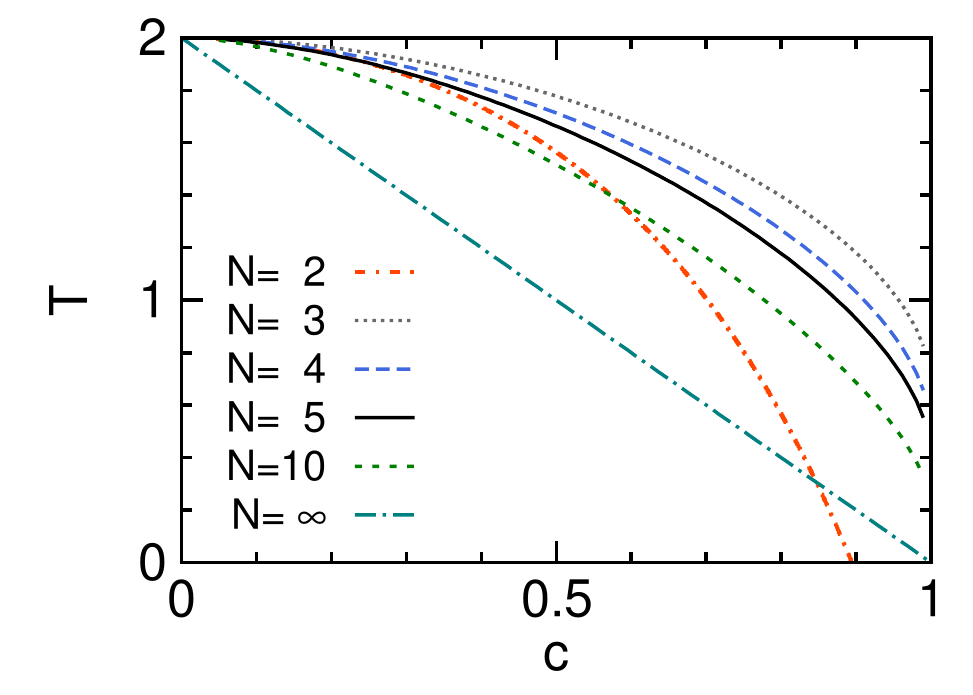}
\caption{
  The critical temperature $T$ at which the permutation-symmetric network becomes locally unstable to perturbations, for symmetric input correlation matrix $\bm{C}$ with $c_{i i}=1$ and non-diagonal elements all being equal to $c$. The results for $N=2, 3, 4, 5, 10$ are obtained by numerical eigenvalue computation on the Hessian matrix (\ref{eq:Lhessian}) and they are in full agreement with analytical predictions [Eq.~(\ref{eq:Tab0})]; the $N=\infty$ is obtained by analytical asymptotic analysis.
}
\label{fig:localstab}
\end{figure}

Figure~\ref{fig:localstab} demonstrates that the curve of $T_2^{\textrm{rb}}$ for $N=2$ is different from the corresponding curves for $N \geq 3$. After a careful examination of the numerically obtained eigenvectors of the Hessian matrices $\bm{L}$ for $N=2$ and $N\geq 3$, we realize that the systems containing $N\geq 3$ units have more ways to break the permutation symmetry and reciprocity than the two-unit system. For example, one additional symmetry for $N\geq 3$ is the rotation symmetry among the $N$ units. Our numerical results suggest that the permutation symmetry of the system starts to be unstable at $T_N^{\textrm{rb}}$ and it may evolve to an optimal solution with rotation symmetry among the $N$ units. This is easy to understand: a rotation-symmetric weight matrix $\bm{W}$ could be regarded as a long wavelength perturbation of a permutation-symmetric one.

Based on this insight, we compute the minimum eigenvalue of the Hessian matrix $\bm{L}$ corresponding to a rotation-symmetric perturbation vector $\{\varepsilon_{i j}\}$. We then obtain an explicit expression of the critical temperature $T_N^{\textrm{rb}}$ for $N \geq 3$ as
\begin{equation}
  T_N^{\textrm{rb}} =
  2 \Bigl( \frac{\sqrt{ 1 + (N - 1) c} + (N-1) \sqrt{1 - c} }{N} \Bigr)^2 \; , 
  \label{eq:Tab0}
\end{equation}
by locating the point at which this eigenvalue changes sign. Another way of deriving this expression will be described in the next section. This analytical formula is in full agreement with our numerical results in Fig.~\ref{fig:localstab}. The $N\rightarrow \infty$ limit of this expression, $T_{N\rightarrow \infty}^{\textrm{rb}} = 2 ( 1- c)$, is also shown in this figure.

\subsection{On the ideal-gas law, cyclic-dominance, and excitation--inhibition balance}

The ideal-gas law $E = \frac{N}{2} T$, possible only for $N\geq 3$, was first observed through numerical computations. It was found to hold true in an extended temperature range immediately after interaction reciprocity (and permutation symmetry) is broken. We now confirm these numerical results through rigorous analytical derivations.

First, as noticed in the main text, the energy and entropy can be computed as
\begin{equation}
  E = \sum_{i=1}^{N}  \epsilon_i \; , \quad \quad 
  S = \frac{1}{2} \sum_{i=1}^{N}\ln {\epsilon_i}  - \frac{1}{2} 
  \ln\bigl( \textrm{Det}(\bm{C} ) \bigr)
  \; ,
\end{equation}
with $\epsilon_1, \ldots, \epsilon_N$ being the $N$ non-negative eigenvalues of the symmetric matrix $\bm{M}$ defined in Eq.~(\ref{eq:matrixM}). By applying the following sum--product inequality
\begin{equation}
  \frac{1}{N} \sum\limits_{i=1}^N \epsilon_i
  \geq \Bigl( \prod_i \epsilon_i \Bigr)^{\frac{1}{N}}
  \; ,
  \label{eq:meanineq}
\end{equation}
we obtain the following upper bound for the entropy,
\begin{equation}
  S \leq \frac{N}{2} \ln \Bigl(\frac{E}{N} \Bigr)
  - \frac{1}{2} \ln \bigl( \textrm{Det} ( \bm{C} ) \bigr) \; .
  \label{eq:Einequality}
\end{equation}
This upper bound of $S$ will be attained only if all the eigenvalues $\epsilon_i$ take the same value $E/N$. If this is the case for a continuous range of $E$, then by applying $T = \frac{ {\rm d} E}{{\rm d} S}$ as required by the minimization of the free energy $F$, we get the ideal-gas law $E = \frac{N}{2} T$.

From the ideal-gas law we can easily derive that
\begin{equation}
  \bm{C} = \frac{T}{2} \bigl( \bm{I} + \bm{W} \bigr)
  \big( \bm{I} + \bm{W} \bigr)^\top \; ,
\end{equation}
which means that the general form of the matrix $\bm{I}+\bm{W}$ is
\begin{equation}
  \bm{I} + \bm{W} \; = \; \sqrt{ \frac{2}{T} } \bm{C}^{1/2} \bm{U}
  \; ,
  \label{eq:MtxUdef}
\end{equation}
where $\bm{U}$ is certain real orthogonal matrix satisfying $\bm{U} \bm{U}^\top = \bm{U}^\top \bm{U} = \bm{I}$, and the square-root of $\bm{C}$ is expressed as
\begin{equation}
  \bm{C}^{1/2} = \sqrt{1 - c}\, \bm{I} + \frac{\bigl( \sqrt{1 + (N-1) c} - \sqrt{1-c} \bigr)}{N}\,
  \begin{bmatrix}
    1
    \\
    1
    \\
    \vdots
    \\ 1
  \end{bmatrix}
  \,
  \begin{bmatrix}
    1 \; & 1 \; & \ldots \; & 1 \;
  \end{bmatrix}
  \; .
\end{equation}
Because the diagonal elements of $\bm{W}$ is identical to zero, we see from Eq.~(\ref{eq:MtxUdef}) that each column of the orthogonal matrix $\bm{U}$ needs to satisfy the additonal condition of
\begin{equation}
  \label{eq:idgcondition}
  \sum\limits_{j=1}^{N} \bigl( \bm{C}^{1/2}\bigr)_{i j} U_{j i}
  \, = \, \sqrt{ \frac{T}{2} }
  \; , \quad \quad (i = 1, \ldots, N ) \; .
\end{equation}
If the temperature $T$ is too large, it is impossible to satisfy these $N$ conditions (\ref{eq:idgcondition}) by any orthogonal matrix $\bm{U}$. When $T$ is lowered to the critical value $T_N^{\textrm{rb}}$ as given by Eq.~(\ref{eq:Tab0}), we find that these conditions are uniquely satisfied by the orthogonal matrix $\bm{U} = \bm{I}$. At this critical value, the matrix $\bm{W}$ is still permutation-symmetric and all its non-diagonal elements are
\begin{equation}
  w_{i j} = \frac{\sqrt{1 + (N-1) c} - \sqrt{1-c} }{ \sqrt{1 + (N-1) c} + (N - 1) \sqrt{1 - c} } \; .
  \label{eq:Wab0}
\end{equation}

For the ideal-gas law to hold at $T < T_N^{\textrm{rb}}$, the constraints (\ref{eq:idgcondition}) require that the $i$-th column vector of $\bm{U}$ should be staying on an $(N-1)$-dimensional ellipsoid on a $N$-dimensional hypersphere of unit radius. These $N$ constraints are compatible with the constraint of $\bm{U}$ being an orthogonal matrix, as long as $N \geq 3$. To demonstrate this most clearly, we consider the type of weight matrices with rotational symmetry.

The general form of such a matrix is
\begin{equation}
  \bm{I} + \bm{W} = \begin{bmatrix}
    w_0  &  w_1  & w_2    & w_3    & \cdots & w_{N-1} \\
    w_{N-1} & w_0      & w_1    & w_2    & \cdots & w_{N-2} \\
    w_{N-2} & w_{N-1} & w_0      & w_1    & \cdots & w_{N-3} \\
    w_{N-3} & w_{N-2} & w_{N-1} & w_0      & \cdots & w_{N-4} \\
    \vdots & \vdots & \vdots & \vdots & \ddots & \vdots \\
    w_1    & w_2    & w_3    & w_4    & \cdots & w_0
  \end{bmatrix}
  \; ,
\end{equation}
where $w_0 \equiv 1$ and $w_1, \ldots, w_{N-1}$ are the other weight parameters. Let us define $N$ discrete frequencies as $\hat{\omega}_j \equiv 2 \pi j / N$ for any $j = 0, \ldots, N-1$ (so $\hat{\omega}_0 = 0,\; \hat{\omega}_1 = 2\pi /N, \; \ldots$). Then it is easy to check that the $j$-th eigenvalue $\lambda_j$ of $\bm{I}+\bm{W}$ is $\lambda_j = \sum_{k=0}^{N-1} w_k e^{i k \hat{\omega}_j}$ for $j=0, \ldots, N-1$, and the corresponding complex eigenvector is $\bm{r}_j = (1, e^{i \hat{\omega}_j}, e^{2 i \hat{\omega}_j}, \ldots, e^{(N-1) i \hat{\omega}_j})^\top /\sqrt{N}$. It is easy to check that $\bm{r}_j^\top \bm{r}_j^* = 1$ and $\bm{r}_i^\top \bm{r}_j^* = 0$ for $j\neq i$. The relationship between the eigenvalues $\{\lambda_j\}$ and the weight parameters $\{w_j\}$ can also be written as
\begin{equation}
  \begin{bmatrix}
    w_0 \\
    w_1 \\
    w_2 \\
    \vdots \\
    w_{N-1}
  \end{bmatrix}
  \; = \frac{1}{N} \;
    \begin{bmatrix}
      1 & 1 & 1 & \ldots & 1 \\
      1 & e^{-i \hat{\omega}_1} & e^{-i \hat{\omega}_2} & \ldots & e^{-i \hat{\omega}_{N-1}} \\
      1 & e^{-2 i \hat{\omega}_1} & e^{-2 i \hat{\omega}_2} & \ldots & e^{-2 i \hat{\omega}_{N-1}} \\
      \vdots & \vdots & \vdots & \ddots & \vdots \\
      1 & e^{- (N-1) i \hat{\omega}_1} & e^{- (N-1) i \hat{\omega}_2} & \ldots & e^{-(N-1) i \hat{\omega}_{N-1}}
    \end{bmatrix}
    \;
    \begin{bmatrix}
      \lambda_0 \\
      \lambda_1 \\
      \lambda_2 \\
      \vdots \\
      \lambda_{N-1}
    \end{bmatrix}
    \; .
    \label{eq:wlmatrix}
\end{equation}

If $N$ is odd so that $N = 1 + 2 N_h$, then one of the eigenvalues ($\lambda_0$) is real, and the remaining $(N-1)$ eigenvalues form $N_h$ complementary pairs, namely $\lambda_{j} = \lambda_{N-j}^*$ for $j=1, \ldots, N_h$. Because of the fact $e^{-i \hat{\omega}_j} = e^{i \hat{\omega}_{N-j}}$, we can check from equation (\ref{eq:wlmatrix}) that all the weight parameters $w_j$ must be real-valued. If $N$ is even with $N=2 N_h$, then two of the eigenvalues ($\lambda_0$ and $\lambda_{N_h}$) are real-valued, while the remaining $N-2$ eigenvalues form $N_h-1$ complementary pairs ($\lambda_j = \lambda_{N-j}^*$ for $j=1, \ldots, N_h-1$). Notice that $e^{-i \hat{\omega}_{N_h}} = -1$. We can also check that all the weight parameters as determined by Eq.~(\ref{eq:wlmatrix}) are again real-valued for $N$ being even.

The mean energy $E$ and the entropy $S$ for this type of matrices are, respectively,
\begin{eqnarray}
  E & = & \frac{1 + (N-1) c}{ |\lambda_0|^2 }
  + \sum\limits_{j = 1}^{N-1} \frac{ 1 - c }{ | \lambda_j |^2 } \; , \\
  S & = & - \sum\limits_{j=0}^{N-1} \ln | \lambda_j | \; .
\end{eqnarray}
Using the sum--product inequality for the eigenvalues $|\lambda_i|$ (see Eq.~(\ref{eq:meanineq})), we obtain that
\begin{equation}
  E \geq N \Bigl[ \bigl(1 + (N-1) c \bigr) (1-c)^{N-1} \Bigr]^{1/N} e^{2 S / N}
  \; .
\end{equation}
The equality holds only if $|\lambda_j|^2$ for all $j= 1, \ldots, N-1$ are equal to each other and that 
\begin{equation}
  \frac{1 + (N-1) c}{| \lambda_0|^2 } = \frac{ 1 - c}{ |\lambda_j |^2 }  \quad \quad (j = 1, \ldots, N-1) \; ,
  \label{eq:l0ljeq}
\end{equation}
for which the temperature is
\begin{equation}
  T = 2 \frac{1+(N-1) c}{ |\lambda_0|^2 } \; ,
  \label{eq:rsT}
\end{equation}
and therefore $E  = \frac{N}{2} T$.

Let us continue to investigate the existence and the degeneracy of the rotation-symmetric matrices satisfying the ideal-gas law. There are multiple constraints on the $N$ eigenvalues $\lambda_i$, including Eq.~(\ref{eq:l0ljeq}), the conjugate property $\lambda_j = \lambda_{N-j}^*$, the condition of
\begin{equation}
  \lambda_0 = N - \sum\limits_{j=1}^{N} \lambda_j
\end{equation}
as inferred from $w_0 \equiv 1$, and furthermore, that the real part of every eigenvalue $\lambda_j$ must be positive to ensure the convergence of the predictive coding dynamics. 

First consider the limiting case of $w_j = w$ for all $j \geq 1$, which corresponds to $\lambda_0 = 1 + (N-1) w$ and $\lambda_j = 1 -w$ for $j\geq 1$. We get from Eqs.~(\ref{eq:l0ljeq}) and (\ref{eq:rsT}) that the only solution is
\begin{eqnarray}
  w & = &  \frac{\sqrt{1 + (N-1) c} - \sqrt{1-c} }{ \sqrt{1 + (N-1) c} + (N - 1) \sqrt{1 - c} } \; ,
  \\
  T & = &  2 \Bigl( \frac{\sqrt{ 1 + (N - 1) c} + (N-1) \sqrt{1 - c} }{N} \Bigr)^2  
  \; .
\end{eqnarray}
This point $(w, T)$ is nothing but the critical point at which the permutation-symmetric solution becomes unstable [see Eqs.~(\ref{eq:Tab0}) and (\ref{eq:Wab0})].

Notice that for $N = 2$ it is impossible to have rotational symmetry when reciprocity is broken ($w_{1 2} \neq w_{2 1}$), so the analysis of this section does not hold. The instability of the permutation symmetry in $N=2$ is not towards rotational symmetry, but towards one unit dominating the other one.

For $N = 3$, let us denote $\lambda_1 = a_1 + i b_1$ and $\lambda_2 = a_1 - i b_1$. Then we have
\begin{subequations}
  \begin{align}
    a_1 & = (3- \lambda_0)/2 \; ,
    \\
    b_1 & = \sqrt{ \frac{1-c}{1 + 2 c} \lambda_0^2 - \frac{(3-\lambda_0)^2}{4} }
    \; ,
    \\
    T & =2 \frac{1+2 c}{ \lambda_0^2 }
    \; .
  \end{align}
\end{subequations}
We see that a unique rotation-symmetric solution exists for $T \in (\frac{2( 1+2 c)}{9}, T_c^{\textrm{rb}})$. There is no non-trivial degeneracy within the rotation-symmetric solutions.

For $N=4$, we have two real eigenvalues $\lambda_0$ and $\lambda_2$ and two complementary complex eigenvalues $\lambda_{1, 3} = a_1 \pm i b_1$. The solution is determined by
\begin{subequations}
  \begin{align}
    \lambda_2 & = \sqrt{ \frac{1-c}{1 + 3 c} } \lambda_0 \; , \\
    a_1 & = \frac{4 - \bigl( 1 +  \sqrt{ \frac{1-c}{1+3 c} } \big) \lambda_0 }{2} \; ,
    \\
    b_1 & = \sqrt{ \frac{1-c}{1 + 3 c} \lambda_0^2 - a_1^2 } \; ,
    \\
    T & =2 \frac{1+3 c}{ \lambda_0^2 }
    \; .
  \end{align}
\end{subequations}
At each valid temperature $T$, there is only one solution of $\lambda_0$, and the values of $\lambda_2$, $a_1$ and $b_1$ are then all uniquely determined. So there is also no non-trivial degeneracy within the rotation-symmetric solutions.

For $N=5$, we have one real-valued eigenvalue $\lambda_0$, and two pairs of complex eigenvalues: $\lambda_{1, 4} = a_1 \pm i b_1$ and $\lambda_{2, 3} = a_2 \pm i b_2$. To eliminate the trivial symmetry we may require that $a_1 \geq a_2$. Then we have
\begin{subequations}
  \begin{align}
    a_1 + a_2 & = \frac{5}{2} - \frac{1}{2} \lambda_0  \; , \\
    a_1^2 + b_1^2 & = \frac{ 1 - c }{1 + 4 c} \lambda_0^2 \; , \\
    a_2^2 + b_2^2 & = \frac{ 1 - c }{1 + 4 c} \lambda_0^2 \; , \\
    T & =2 \frac{1+4 c}{ \lambda_0^2 }
    \; .
  \end{align}
\end{subequations}
Given the temperature $T$, the value of $\lambda_0$ is uniquely determined. The sum $a_1 + a_2$ is then constrained to a fixed value but the individual values of $a_1$ and $a_2$ are not strictly constrained. Therefore there is an infinite degrees of degeneracy within the rotation-symmetric solutions. This same conclusion holds for all $N \geq 5$.

Rotation-symmetric weight matrices are special examples of more general cyclic-dominant (CD) matrices. The concept of cyclic-dominance is very easy to understand intuitively. Roughly speaking, it means one unit (say $p_1$) of the system is most strongly inhibited by another unit (say $p_2$) while $p_1$ only weakly affects $p_2$, and unit $p_2$ is most strongly inhibited by yet another unit $p_3$ while $p_2$ only weakly affects $p_3$, ..., and finally unit $p_N$ is most strongly inhibited by unit $p_1$ while $p_N$ only weakly affects $p_1$. The directed cycle $p_1 \leftarrow p_2 \leftarrow p_3 \leftarrow \ldots \leftarrow p_N \leftarrow p_1$ is a global closed loop of dominance. It is relatively easy to quantitatively measure the degree of cyclic-dominance in the system. For example, we may design an order parameter $\mathcal{O}_{CD}$ as
\begin{equation}
  \mathcal{O}_{CD} = 1 - \min\limits_{(p_1,\ p_2,\ p_N)} \biggl|
  \frac{ \sum_{i=1}^{N-1} w_{p_{i+1} \ p_i}  }{
     \sum_{i = 1}^{N-1} w_{p_i\ p_{i+1} } } \biggr|
  \; ,
\end{equation}
where $(p_1, p_2, \ldots, p_N)$ denotes a permutation of the indices of the $N$ units. We see that $\mathcal{O}_{CD}$ compares the most dominant cycle of directed interactions among $N$ units and the reverse directed cycle of interactions. If there is no or only very weak cyclic dominance then $\mathcal{O}_{CD} \approx 0$; if there is strong cyclic dominance (the limiting case being the $N$ units form a single cycle of one-way interactions but no reverse interactions), then $\mathcal{O}_{CD}$ will take a value close to unity. The order parameter $\mathcal{O}_{CD}$ for the five units shown in the right panel of Fig.~2(b) of the main text is $0.996$. 

Besides rotation-symmetric solutions there are other types of matrices, such as cyclic-dominant matrices without the property of rotational symmetry and different forms of EI-balanced networks (some concrete examples are given in the next sections). This will bring in additional degrees of degeneracy within the optimal solution space, further increasing the flexibility of achieving the ideal-gas law.

The units in a cyclic-dominant optimal weight matrices $\bm{W}$ may be regarded as forming only one component in the global level. We also encounter EI-balanced optimal weight matrices in our numerical computations. Roughly speaking, an EI-balanced weight matrix $\bm{W}$ are formed by two components of units. Group $g_I$ contains inhibitory units $i$ whose output synaptic weights $w_{k i}$ to other units $k$ are mostly inhibitory (i.e., $w_{k i} > 0$); the other group $g_E$ contains excitatory units $j$ whose output synaptic weights $w_{k j}$ to other units $k$ are mostly excitatory (i.e., $w_{k j} < 0$). An example of EI-balanced optimal matrix is shown in Fig.~2(b) of the main text. 

One quantitative measure of EI-balance is to compute the net synaptic weights on all the $N$ units,
\begin{equation}
  \mathcal{O}_{EI} = \frac{1}{N} \sum\limits_{i = 1}^{N} \biggl[ 1 -  \frac{ \bigl| \sum_{j \neq i} w_{i j} \bigr|}{
      \sum_{j \neq i} |w_{i j} |} \biggr] \; .
\end{equation}
Notice that if the sum of input weights to unit $i$ is $\sum_{j \neq i} w_{i j} \approx 0$, then this unit is EI-balanced and it contributes a value $\approx 1$ to the order parameter $\mathcal{O}_{EI}$; while if all the input weights to unit $i$ are of the same sign, then it contributes a term $0$ to the order parameter. The EI-balanced optimal matrix of Fig.~2(b) has a value of $\mathcal{O}_{EI} = 0.626$.

Another more coarse-grained measure of EI-balance is simply to compare the number $M_+$ of postive synaptic weights and the number $M_-$ of negative synaptic weights:
\begin{equation}
  1 - \frac{\bigl| M_+ - M_- \bigr|}{ M_+ + M_-} \; .
\end{equation}
The EI-balanced matrix of Fig.~2(b) attains the maximum value $1.0$ in terms of this order parameter.

\subsection{Numerical optimization methods}

Analytical solutions of the free energy minimization problem are hard to achieve for systems containing $N\geq 3$ units, especially at low temperatures. We resort to numerical optimization methods to solve this problem. Two simulated annealing search processes are implemented to construct optimal weight matrices. One process runs at a fixed temperature value $T$ and aims at reaching the global minimum of the free energy $F$. The other works at a fixed entropy value $S$ and attempts to reach the minimum energy $E$ that is compatible with this entropy.

We introduce an annealing parameter $\kappa$ and assign it a small initial value (say $\kappa = 10^2$). Then we run $n_0$ (e.g., $n_0= 10^5$) stochastic trials to update $\bm{W}$ at each epoch of the optimization process.  After one such epoch of trials, the next epoch then follows with a slightly increased value of $\kappa$ (such as $ \kappa \leftarrow 1.01 \times \kappa$), initializing $\bm{W}$ by the so-far reached best weight matrix in this process. This annealing process is continued for hundreds of epochs until $\kappa$ reaches a high value (say $10^{8}$). The tentative optimal weight matrix is then further refined (if possible) by running greedy descent at $\kappa = \infty$ for $10 \times n_0$ trials, and the final matrix $\bm{W}$ is returned as one candidate optimal solution. To make sure that the global minimum of free energy (at fixed $T$) or energy (at fixed $S$) has been achieved, we repeat this whole $\kappa$-annealing process a number of times starting from different initial conditions and random number seeds and check whether the outcomes have converged to the same minimum value. 

\subsubsection*{Temperature $T$ being fixed}

If the temperature $T$ is fixed during the whole simulation process, in each elementary trial we propose a new weight matrix $\bm{W}^\prime$ based on the current weight matrix $\bm{W}$ according to the following updating rules:
\begin{enumerate}
\item[(a):] Single-element perturbation. The current weight matrix $\bm{W}$ is copied to the new matrix $\bm{W}^\prime$ but with a single randomly chosen element $w_{i j}$ being changed to $w_{i j}^\prime = w_{i j} + \delta_{i j}$. The deviation $\delta_{i j}$ is uniformly distributed in $(- \Delta_1, \Delta_1)$.
\item[(b):] All-elements perturbation. The current weight matrix $\bm{W}$ is copied to the new matrix $\bm{W}^\prime$ but with every element $w_{i j}$ being changed to $w_{i j}^\prime = w_{i j} + \delta_{i j}$. The deviation $\delta_{i j}$ is an independently and uniformly distributed random number in the range $(- \Delta_2, \Delta_2)$.
\end{enumerate}
The eigenvalues of the matrix $\bm{I}+\bm{W}^\prime$ are first computed. If the real part of an eigenvalue is below a bottomline value (fixed to be $10^{-5}$ in this work), then $\bm{W}^\prime$ is considered as invalid and the proposed update is rejected. If the eigenvalue bottomline constraint is satisfied by all the eigenvalues, then the free energy difference $\delta F$ induced by the proposed weight matrix $\bm{W}^\prime$ is computed. If $\delta F \leq 0$ then $\bm{W}^\prime$ is copied back to $\bm{W}$ (the proposed update being accepted). If $\delta F > 0$, then the proposed $\bm{W}^\prime$ is accepted only with probability $e^{-\kappa \delta F}$ ($\bm{W}$ is replaced by $\bm{W}^\prime$) and it is rejected with the remaining probability $1 - e^{-\kappa \delta F}$ (the old $\bm{W}$ is kept).

We choose single-element and all-elements perturbations with equal probability $1/2$ at each elementary trial. To improve the search efficiency, both $\Delta_1$ and $\Delta_2$ are dynamically adjusted so that the success rate of weight matrix change will be approximately $0.5$. We also set the lower bound $10^{-6}$ to the values of $\Delta_1$ and $\Delta_2$ to avoid the simulation process being trapped in a local minimum. If this lower bound is reach by (say) $\Delta_1$ in one epoch, we reset its value to $\Delta_1 = 10^{-4}$ in the next epoch to further increase the chance of escaping a local minimum of $F$.

\subsubsection*{Entropy $S$ being fixed}

If the whole simulation process is run under the constraint of fixed entropy, then in each elementary trial a new weight matrix $\bm{W}^\prime$ is generated by modifying a single row or a single column of the old matrix $\bm{W}$. Consider a row modification which generates the $i$-th row of $\bm{W}^\prime$ as
\begin{equation}
  w_{i k}^\prime  =   w_{i k} + \delta_{i k} \; ,
  \quad \quad \quad (k = 1, \ldots, N) \; .
\end{equation}
The constraint of fixed entropy $S$ means the following linear constraint on the $N-1$ perturbations $\delta_{i k}$:
\begin{equation}
  \label{eq:rowconst}
  \sum\limits_{k \neq i} \delta_{i k} \Bigl( \frac{\bm{I}}{\bm{I}+\bm{W}} \Bigr)_{k i} = 0 \; .
\end{equation}
We can first generate $N-1$ independent deviations $\delta_{i k}$ by sampling in the range $(- \Delta_1, \Delta_1)$, and then add a correction term $\delta_0$ to all these deviations to satisfy the constraint (\ref{eq:rowconst}).

Similarly, if a column $j$ of the new matrix $\bm{W}^\prime$ differs from the old matrix $\bm{W}$ by $w_{k j}^\prime = w_{k j} + \delta_{k j}$, then the deviations $\delta_{k j}$ shall satisfy the following constraint:
\begin{equation}
  \label{eq:columnconst}
  \sum\limits_{k \neq j} \Bigl( \frac{\bm{I}}{\bm{I}+\bm{W}} \Bigr)_{j k} \delta_{k j} = 0 \; .
\end{equation}

After a new matrix $\bm{W}^\prime$ has been proposed by row modification (with probability one-half) or by column modification (again with probability one-half), the eigenvalues of the matrix $\bm{I}+\bm{W}^\prime$ are first computed. If the real part of an eigenvalue is below the bottomline value ($10^{-5}$), then the proposed update is rejected. If the eigenvalue bottomline constraint is satisfied by all the eigenvalues, then the energy difference $\delta E$ induced by the proposed weight matrix $\bm{W}^\prime$ is computed. If $\delta E \leq 0$ then $\bm{W}$ is replaced by $\bm{W}^\prime$. If $\delta E > 0$, then the proposed $\bm{W}^\prime$ is accepted with probability $e^{-\kappa \delta E}$ and it is rejected with the remaining probability $1 - e^{-\kappa \delta E}$.

\begin{figure}[h]
  \centering
  \includegraphics[angle=270,width=0.5\textwidth]{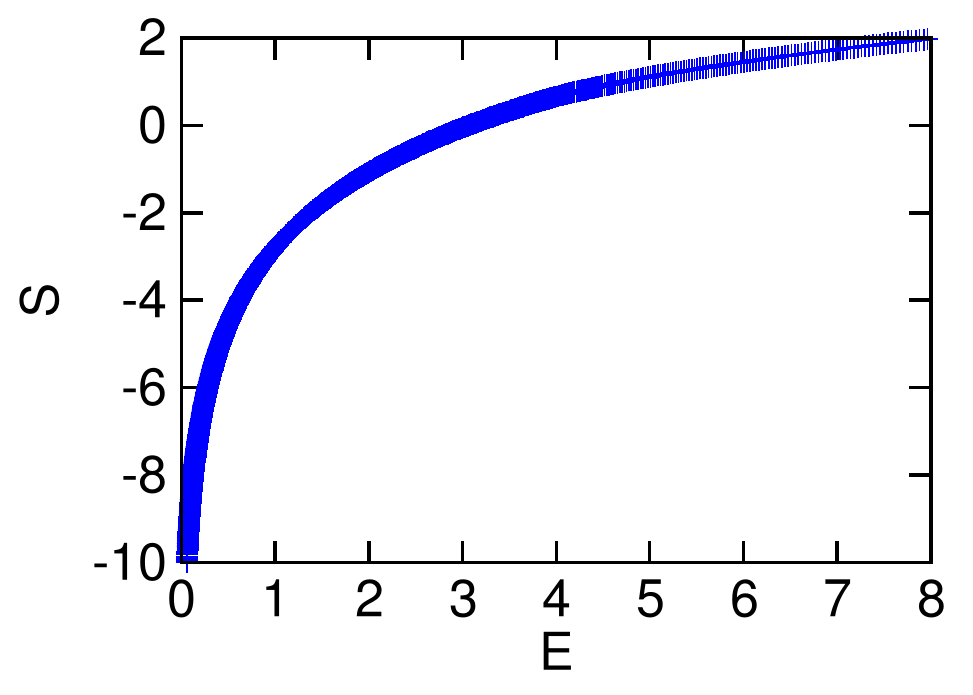}
  \caption{
    The numerically obtained relationship between entropy $S$ and energy $E$ for $N=5$ and $c=0.6$. 
  }
  \label{fig:Sn5c0p6}
\end{figure}

\subsection{Some example matrices for $N=5$}

The phase diagram for $N=5$ is shown in Fig.~3 of the main text. Here we offer some additional concrete samples of optimal weight matrices to help appreciate this phase diagram. The different phase regions are discussed by decreasing the temperature $T$ along a vertical line of fixed pair correlation $c$. Here we mainly consider $c = 0.6$, for which the relationship between $E$ and $S$ is shown in  Fig.~\ref{fig:Sn5c0p6}.

\subsubsection*{Phase region $\alpha$}

For example, at the point $c=0.6$ and $T=1.6507$ (slightly above the critical temperature $1.5305$), the permutation-symmetric matrix is
\begin{equation}
  \alpha: \quad
  \begin{bmatrix}
    0      &   0.2900 &  0.2900 &  0.2900  & 0.2900 \\
    0.2900 &   0      &  0.2900 &  0.2900  & 0.2900 \\
    0.2900 &   0.2900 &  0      &  0.2900  & 0.2900 \\
    0.2900 &   0.2900 &  0.2900 &  0       & 0.2900 \\
    0.2900 &   0.2900 &  0.2900 &  0.2900  & 0
  \end{bmatrix}
  \; ,
\end{equation}
with entropy $S = 0.6$ and energy $E = 3.9030$, and the eigenvalues of $\bm{I}+\bm{W}$ being $\{ 2.1602, \; 0.7100\pm 8 \times 10^{-7},\; 0.7100 \pm 2\times 10^{-7}\}$.

\subsubsection*{Phase region $\beta_1$}

We show two degenerate optimal matrices here for the point $c=0.6$, $T=0.45152$ 
in phase region $\beta_1$:
\begin{equation}
  \begin{bmatrix}
    0       & -0.477 & 0.909 & 0.852 & 1.285 \\
    1.346  & 0       & 0.059 & 1.114 & 0.612 \\
    0.335  & 1.123  & 0     & 0.136 & 1.428 \\
    0.622  & 0.614  & 1.633 & 0     & -0.006 \\
    -0.277 & 0.518  & 0.513 & 1.679 & 0
  \end{bmatrix}
  \; , \quad
  \begin{bmatrix}
    0      & -0.399 & 1.172 & 1.049 & 0.892 \\
    -0.220 & 0      & 1.411 & 0.916 & 0.743 \\
    -0.639 & -0.550 & 0     & 1.635 & 0.214 \\
    -0.400 & -0.119 & 0.322 & 0     & 1.775 \\
    -0.509 & -0.615 & 1.670 & 0.068 & 0
  \end{bmatrix}
  \; .
\end{equation}
These two solutions have the same entropy $S = -2.50$ and the same energy $E = 1.1288$. The left one is a one-component system with cyclic dominant (CD) interactions but there is no rotation symmetry or other obvious symmetry, and the eigenvalues of $\bm{I}+\bm{W}$ are $\{3.8092,\; 0.5785 \pm 1.2062 i,\; 0.0169 \pm 1.3367 i\}$. The right network is an EI-balanced two-component system with group $g_E$ containing units $1$ and $2$ and group $g_I$ containing the remaining three units, and the eigenvalues are $\{1.3427, \; 1.8103 \pm 1.3213 i,\; 0.0184 \pm 1.3439 i\}$. There is no permutation symmetry within $g_E$, and there is cyclic dominance but no rotation symmetry in $g_I$.

\subsubsection*{Phase range $\beta_2$}

For $c=0.6$, the EI-balanced solution mentioned above reaches the eigenvalue bottomline at a critical temperature value $0.3872$ and its free energy starts to be higher than that of the CD solution. The CD solution will reach the eigenvalue bottomline at a much lower critical temperature $T= 0.2720$, at which its energy susceptibility drops from $5/2$ to $2.0$. This transition is completely due to eigenvalue constraint; there is no any qualitative change in the symmetry of the solution. Here is an example matrix in this phase, obtained at $c=0.6$ and $T = 0.2303$:
\begin{equation}
  \beta_2: \quad
  \begin{bmatrix}
    0       & 0.0249 &  1.0620 &  0.6176  & 2.2955  \\
    1.9752  & 0      & -0.1972 &  1.5409  & 0.6802 \\
    0.9381  & 2.1973 &  0      & -0.1579  & 1.0228 \\
    1.3823  & 0.4586 &  2.1582 &  0       & 0.0004 \\
    -0.2952 & 1.3194 &  0.9771 &  1.9994  & 0  
  \end{bmatrix}
  \; ,
\end{equation}
whose entropy is $S = -4.1$ and energy is $E = 0.56958$ (the eigenvalues of $\bm{I}+\bm{W}$ are $\{ 4.99996,\; 10^{-5} \pm 1.8639 i ,\; 10^{-5} \pm 1.8637 i \}$. This is a CD matrix with all the complex eigenvalues touching the bottomline.

\subsubsection*{Phase region $\gamma_2$}

At $c=0.6$, although the EI solutions are suppressed by the CD solutions in the phase space region $\beta_2$, this suppression will be reversed at a critical temperature $T = 0.2195$, and a discontinuous transition between the CD and EI solutions occurs here. The EI solutions at the lower-temperature phase  region $\gamma_2$ appears to be permutation-symmetric within group $g_E$ and cyclic-dominant with rotation symmetry within group $g_I$. For example, the EI solution at $T = 0.2138$ (immediately after the discontinuous transition) is
\begin{equation}
  \gamma_2 : \quad
  \begin{bmatrix}
    0      &  -1.010 &  1.481 &  1.477  &  1.483 \\
    -1.008 &   0     &  1.481 &  1.479  &  1.484 \\
    -1.165 &  -1.164 &  0      & -0.119  &  2.115 \\
    -1.162 &  -1.160 &  2.120 &  0       & -0.116 \\
    -1.164 &  -1.164 & -0.115 &  2.116  &  0
  \end{bmatrix}
  \; ,
\end{equation}
with entropy $S = -4.35$ and $E = 0.5414$. The eigenvalues of $\bm{I}+\bm{W}$ are $\{2.0086, 1.4957 \pm 2.8409 i, 10^{-5} \pm 1.9345 i \}$.

\subsubsection*{Phase region $\gamma_3$}

As the temperature further decreases to another critical value $T = 0.09505$ (with $c=0.6$), another discontinuous transition occurs to the optimal EI-balanced solution, caused by the breaking of internal rotation symmetry among the three units of group $g_I$. This transition is associated with a small drop of the entropy from $-6.410$ to $-6.417$ and a small drop of the mean energy from $0.2397$ to $0.2390$. To demonstrate the different symmetries of these two types of EI-balanced solutions, we show the two corresponding solutions at the same entropy value $S = -6.414$ and the same energy value $E = 0.2394$:
\begin{equation}
\begin{bmatrix}
  0      & -2.151 & 2.077  & 2.068  & 2.082 \\
  -2.151 & 0      & 2.076  & 2.069  & 2.082 \\
  -2.124 & -2.124 & 0      & -0.684 & 2.673 \\
  -2.119 & -2.119 & 2.683  &  0     & -0.673 \\
  -2.124 & -2.125 & -0.672 &  2.673 & 0
\end{bmatrix}
\; , \quad
\begin{bmatrix}
  0      & -2.139 & 1.503  & 1.708  & 2.821 \\
  -2.139 & 0      & 1.503  & 1.708  & 2.821 \\
  -1.977 & -1.977 & 0      & -1.277 & 2.749 \\
  -1.822 & -1.822 & 3.154  & 0      & 0.172 \\
  -2.436 & -2.436 & -0.618 & 2.102  & 0
\end{bmatrix}
\; .
\end{equation}
Notice that the rotation-symmetry breaking of the group $g_I$ does not involve all the five units of the system; it is an abrupt change only in the subnetwork $g_I$ of the system. Notice also that the permutation symmetry between the two units $1$ and $2$ of group $g_E$ is intact in the EI-balanced phase space region $\gamma_3$.

\subsubsection*{Phase region $\gamma_4$}

As the temperature $T$ further decreases, the EI-balanced optimal solutions of the phase region $\gamma_3$ will reach the eigenvalue bottomline, leading to a kink in the mean energy. This continuous transition occurs at $T = 0.0383$ for $c=0.6$ (with entropy $S \approx -8.73$). Notice that there is no any qualitative change in the symmetry of the EI-balanced matrix, only that all the four complex eigenvalues touch the bottomline.

As the temperature decreases even further, the permutation symmetry between the two units of group $g_E$ will also break. This breaking of internal symmetry will be associated with a very weak continuous transition and a tiny kink in the mean energy. At very low temperatures, then there are no internal symmetries within both group $g_E$ and $g_I$. To give one such example, we show the matrix at a very low temperature $T = 0.02652$ with entropy $S = -9.47$ and mean energy $E = 0.071037$ here:
\begin{equation}
  \gamma_4: \quad
  \begin{bmatrix}
    0      & -3.941 & 1.872  &  0.677  & 6.318 \\
    -4.059 &  0     & 1.912  &  0.652  & 6.259 \\
    -5.355 & -5.293 & 0      & -1.160  & 2.269  \\
    -3.506 & -3.409 & 4.841  &  0      & -4.180 \\
    -3.129 & -3.073 & 2.731  &  6.015  & 0    
  \end{bmatrix}
  \; .
\end{equation}
The eigenvalues of $\bm{I}+\bm{W}$ are $\{4.99996, 10^{-5} \pm 7.1819 i, 10^{-5} \pm 7.0902 i\}$. Notice that cyclic dominance still exist in group $g_I$ and the three units of this group are all different.

\subsubsection*{Phase region $\gamma_1$}

Here we show an optimal solution obtained at $c=0.4$ and $T = 0.0978$ in this second ideal-gas region $\gamma_1$:
\begin{equation}
  \begin{bmatrix}
    0       & -2.0064 &  3.2335 &  0.7069 &  2.1120 \\
    -2.4388 &  0      &  2.7901 &  2.1816 &  0.9742 \\
    -3.1494 & -2.4572 &  0      & -0.6985 &  1.7311 \\
    -1.2998 & -2.4212 &  2.6504 &  0      & -2.2057 \\
    -0.9509 & -2.4603 & -0.1289 &  3.5309 &  0   
  \end{bmatrix}
  \; .
\end{equation}
The entropy is $S = -7$ and energy is $E = 0.2446$. The eigenvalues of $\bm{I}+\bm{W}$ are $\{3.5236, 0.6740 \pm 4.9114 i, 0.0642 \pm 3.5581 i\}$. This is an EI-balanced network with two groups of units. This is no permutation symmetry between the two units $1$ and $2$ of group $g_E$, but there is cyclic dominance within the three different units ($3, 4, 5$) of group $g_I$.

\subsubsection*{Phase region $\delta$}

This phase region is located at $c$ close to unity and $T$ relatively small. There is a discontinuous phase transition between the $\beta_2$ phase and the $\delta$ phase. For example at $c=0.99$ this occurs at $T= 0.0612$. The optimal weight matrix at temperature $T = 0.0613$ (slightly above the critical temperature) is
\begin{equation}
  \beta_2: \quad
  \begin{bmatrix}
    0      & 0.6546 & 1.3549 & 0.9062 & 1.0842 \\
    1.3454 & 0      & 0.7256 & 0.7858 & 1.1431 \\
    0.6451 & 1.2745 & 0      & 0.8718 & 1.2084 \\
    1.0936 & 1.2142 & 1.1282 & 0      & 0.5641 \\
    0.9157 & 0.8569 & 0.7919 & 1.4355 & 0
  \end{bmatrix}
  \; ,
\end{equation}
which has entropy $S = 0.631$ and energy $E = 0.3210$ (the eigenvalues of the matrix $\bm{I}+\bm{W}$ are $\{4.99996, 10^{-5} \pm 0.5712 i, 10^{-5} \pm 0.5711 i\}$. It is a one-component cyclic-dominant network. Immediately after the discontinuous transition, the optimal matrix at $T = 0.0608$ is
\begin{equation}
  \delta: \quad
  \begin{bmatrix}
    0       & 0.1897  & 0.9302 & 0.6233 & 4.7956 \\
    -0.0505 & 0       & 0.9020 & 1.5790 & 3.3432 \\
    -0.5581 & 0.9506  & 0      & 0.9995 & 3.4969 \\
    0.2748  & 0.7426  & 1.5723 & 0      & 3.6772 \\
    -2.7552 & 0.8498  & 1.2921 & 2.8092 & 0
  \end{bmatrix}
  \; ,
\end{equation}
whose entropy is $S = -1.871$ and energy $E = 0.1679$ (the eigenvalues of $\bm{I}+\bm{W}$ are $\{4.99996, \; 10^{-5} \pm 1.0680 i, \; 10^{-5} \pm 1.0672 i\}$). Notice that unit $1$ excites units $2, 3, 5$ but inhibits unit $4$, while the remaining four units are all inhibitory units.

At a lower critical temperature value $T= 0.03445$ there is again a discontinuous phase transition, from an optimal network of the $\delta$ phase to an optimal network of the $\gamma_3$ phase. At the phase transition point, the optimal network of the $\delta$ phase is
\begin{equation}
  \delta: \quad 
  \begin{bmatrix}
    0      & -0.0810 &  0.9556 & 0.5137 & 7.1586 \\
    -0.6198 &  0      &  0.9467 & 1.9758 & 4.9676 \\
    -1.2632 &  0.8165 &  0      & 1.1826 & 5.1382 \\
    1.8054 &  0.6487 &  0.0340 & 0      & 5.7823 \\
    -4.5227 &  0.6593 &  1.4476 & 4.0582 & 0
  \end{bmatrix}
  \; ,
\end{equation}
and its entropy is $S= -3.50$ and energy is $E=0.0914$ (the eigenvalues of $\bm{I}+\bm{W}$ are $\{4.99996,\; 10^{-5} \pm 1.6044 i , \; 10^{-5} \pm 1.6041 i \}$. The optimal network after this transition is
\begin{equation}
  \gamma_3: \quad
  \begin{bmatrix}
    0       & 0.8032 & 0.8644 & 7.5865 & 0.0935 \\
    -2.7520 & 0      & 0.2967 & 4.4990 & 0.1447 \\
    -2.4812 & 1.7006 & 0      & 4.3312 & 0.4106 \\
    -6.4603 & 0.8900 & 1.5595 & 0      & 0.1387 \\
    -2.6629 & 0.8229 & 0.9064 & 4.3890 & 0
  \end{bmatrix}
  \; ,
\end{equation}
whose entropy is $S = -3.6$ and energy is $E = 0.08796$ (the eigenvalues of $\bm{I}+\bm{W}$ are $\{0.8492, \; 2.0754 \pm 6.5191 i , \; 10^{-5} \pm 0.9596 i\}$.

\subsection{Ideal-gas law and phase diagram for the three-unit system ($N = 3$)}

\subsubsection*{The ideal-gas law for general input correlation matrix}

First, it is instructive to study the $N=3$ system with symmetric input with a single parameter $c$ for the pair correlations by exhaustive search. To achieve the ideal-gas law, the following condition must be satisfied by the synaptic weight matrix
\begin{equation}
    \bigl( \bm{I} + \bm{W} \bigr) \; 
    \bigl( \bm{I} + \bm{W} \bigr)^\top = \frac{2}{T} \bm{C} \; .
\end{equation}
We consider the following general form of $\bm{I}+\bm{W}$ as
\begin{equation}
    \bm{I} + \bm{W} = 
    \begin{bmatrix}
        1 & c_2 & c_1 \\
        b_1 & 1 & a_1 \\
        b_2 & a_2 & 1
    \end{bmatrix}
    \; .
\end{equation}
We may require $c_1 \leq c_2$ to eliminate the trivial degeneracy of the synaptic weight matrix. Then we obtain that
\begin{equation}
     \bigl( \bm{I} + \bm{W} \bigr) \; 
    \bigl( \bm{I} + \bm{W} \bigr)^\top \; = \; 
    \begin{bmatrix}
    1 + c_1^2 + c_2^2 & \; b_1 + c_2 + a_1 c_1 & \; b_2 + c_1 + a_2 c_2 \\
    b_1 + c_2 + a_1 c_1 & \; 1 + a_1^2 + b_1^2 & \; a_1 + a_2 + b_1 b_2 \\
    b_2 + c_1 + a_2 c_2 & \; a_1 + a_2 + b_1 b_2 & \; 1 + a_2^2 + b_2^2 
    \end{bmatrix}
    \; .
\end{equation}

To determine the weight parameters, we express
\begin{equation}
    c_1 = x \cos \theta \;, \quad
    c_2 = x \sin \theta
    \label{eq:c1c2}
\end{equation}
with $\theta \in [\frac{\pi}{4}, \frac{5 \pi}{4}]$, and $x = \sqrt{\frac{2}{T} - 1} \geq 0$ (so $T\leq 2$). Then we find that
\begin{subequations}
  \begin{align}
    a_1 & = \frac{(y - c_2) c_1 \pm \sqrt{(1+c_1^2) x^2 - (y - c_2)^2}}{1+c_1^2} \; , \\
    b_1 & = \frac{ y- c_2 \mp c_1 \sqrt{(1+c_1^2) x^2 - (y - c_2)^2}}{1+c_1^2} \; ,
  \end{align}
\end{subequations}
where $y = \frac{2 c}{T}$. Similarly, the expressions for $a_2$ and $b_2$ are
\begin{subequations}
  \begin{align}
    a_2 &= \frac{(y - c_1) c_2 \pm \sqrt{(1+c_2^2) x^2 - (y - c_1)^2}}{1+c_2^2} \; , \\
    b_2 &= \frac{ y- c_1 \mp c_2 \sqrt{(1+c_2^2) x^2 - (y - c_1)^2}}{1+c_2^2} \; .
  \end{align}
\end{subequations}
Notice that the following condition must be satisfied by $a_1, b_1, a_2, b_2$:
\begin{equation}
    a_1 + a_2 + b_1 b_2 = y \; ,
\end{equation}
which means that
\begin{subequations}
  \begin{align}
    a_2 & = \frac{ y- a_1 \mp b_1 \sqrt{(1+b_1^2) x^2 - (y - a_1)^2}}{1+b_1^2} \; , \\
    b_2 & = \frac{(y - a_1) b_1 \pm \sqrt{(1+b_1^2) x^2 - (y - a_1)^2}}{1+b_1^2} \; .
  \end{align}
\end{subequations}
The other constraints (to ensure the weight values being real) are
\begin{equation}
(1+ c_1^2) x^2 - (y-c_2)^2  \geq 0 \; , \quad
(1+ c_2^2) x^2 - (y - c_1 )^2 \geq 0 \; .
\end{equation}
All the eigenvalues of $\bm{I}+\bm{W}$ should have positive real part. 

At each value of $\theta$ there are up to four distinct solutions of $(a_1, a_2, b_1, b_2)$. They can be determined accurately by numerical computations. The numerical results are the following.

For the EI-balanced network, numerical results indicate that the upper-bound temperature can be obtained by setting $c_1 = c_2 = - \frac{x}{\sqrt{2}}$. Then $b_1 = b_2$ and $a_1=a_2$. We can then obtain the upper-bound temperature. For example, the upper-bound EI-balanced temperature at $c=0.35$ is $T=0.264684$, and the corresponding network is
\begin{equation}
    \bm{W} = \begin{bmatrix}
        0 & -1.81055 &	-1.81055 \\
2.25323	& 0 &	-1.2162 \\
2.25323	& -1.2162	& 0 
    \end{bmatrix} \; .
\end{equation}

%

This EI-balanced system breaks the internal permutation symmetry within group $g_E$ (units $2$ and $3$) to achieve the ideal-gas law. On the other hand, the lower-bound temperature for the ideal-gas law for the EI-balanced network can be computed numerically by computing the eigenvalues of $\bm{W}$. The temperature upper-bound and lower-bound are shown in Fig.~\ref{fig:EIN3}.

\begin{figure}[h]
    \centering
    \subfigure[]{
    \includegraphics[angle=270,width=0.45\textwidth]{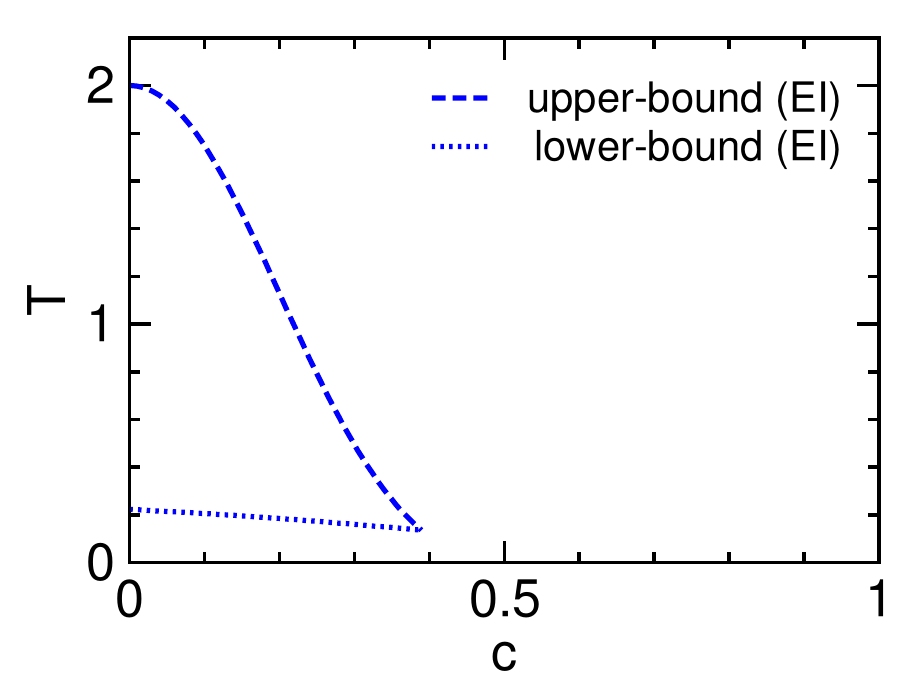}
    \label{fig:EIN3}
    }
  \subfigure[]{
    \includegraphics[angle=270,width=0.45\textwidth]{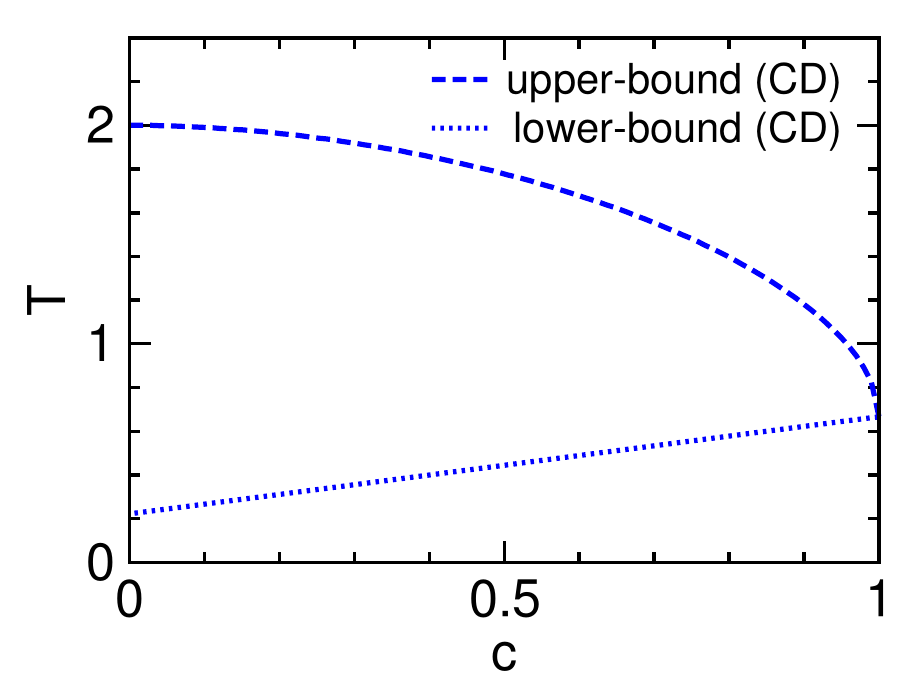}
    \label{fig:CDN3}
    }
  \caption{(a) The upper-bound and lower-bound temperatures for the ideal-gas law of EI-balanced optimal solutions (a) and for cyclic-dominant optimal solutions (b). Number of units is $N=3$, and $c$ is the input correlation parameter.
  }
  \label{fig:N3c}
\end{figure}

The other types of network realizing the ideal-gas law can all be obtained. For $N=3$ these are cyclic-dominant networks.  The temperature upper-bound and lower-bound for these ideal-gas CD networks are shown in Fig.~\ref{fig:CDN3}. By comparing Figs.~\ref{fig:EIN3} and \ref{fig:CDN3} we see that there is a region in the phase space in which EI-balanced and CD optimal solutions coexist.

The above analysis also works for the case of $\bm{C}$ having different values:
\begin{equation}
    \bm{C} = \begin{bmatrix}
        1 & c_{1 2} & c_{1 3} \\
        c_{1 2} & c_{2 2} & c_{2 3} \\
        c_{1 3} & c_{2 3} & c_{3 3}
    \end{bmatrix}
    \; ,
    \label{eq:iglcondition}
\end{equation}
with parameters $c_{2 2}$, $c_{3 3}$, $c_{1 2}$, $c_{1 3}$, and $c_{2 3}$ (the value of $c_{1 1}$ is fixed to be $c_{1 1} = 1$ and we require that $1\leq c_{2 2} \leq c_{3 3}$).  Then we still have Eq.~(\ref{eq:c1c2}) for the weight parameters $c_1$ and $c_2$, and the equations for $a_1$ and $b_1$ are
\begin{subequations}
  \begin{align}
    a_1 &= \frac{(y_1 - c_2) c_1 \pm \sqrt{(1+c_1^2) x_2^2 - (y_1 - c_2)^2}}{1+c_1^2} \; , \\
    b_1 & = \frac{ y_1- c_2 \mp c_1 \sqrt{(1+c_1^2) x_2^2 - (y_1 - c_2)^2}}{1+c_1^2} \; ,
  \end{align}
\end{subequations}
where $x_2 = \sqrt{ \frac{2 c_{2 2}}{T} - 1}$ and $y_1 = \frac{2 c_{1 2}}{T}$. Similarly, the expressions for $a_2$ and $b_2$ are
\begin{subequations}
  \begin{align}
    a_2 & = \frac{(y_2 - c_1) c_2 \pm \sqrt{(1+c_2^2) x_3^2 - (y_2 - c_1)^2}}{1+c_2^2} \; , \\
    b_2 & = \frac{ y_2 - c_1 \mp c_2 \sqrt{(1+c_2^2) x_3^2 - (y_2 - c_1)^2}}{1+c_2^2} \; .
  \end{align}
\end{subequations}
where $x_3 = \sqrt{ \frac{2 c_{3 3}}{T} - 1}$ and $y_2 = \frac{2 c_{1 3}}{T}$. The other constraint is
\begin{equation}
    a_1 + a_2 + b_1 b_2 = y_3 \; ,
\end{equation}
where $y_3 = \frac{2 c_{2 3}}{T}$.

We now examine a particular example, with correlation matrix
\begin{equation}
    \bm{C} = \begin{bmatrix}
        1 & 0.15 & 0.1 \\
        0.15 & 1.1 & 0.25 \\
        0.1 & 0.25 & 1.2 
    \end{bmatrix}
    \; .
    \label{eq:Cgeneral}
\end{equation}
Examining all the possible solutions of the ideal-gas-law condition (\ref{eq:iglcondition}), we find that the ideal-gas law is achievable for temperatures $T \in [0.19402, 1.98262]$. At the maximum temperature $T = 1.98262$ the ideal-gas matrix is
\begin{equation}
    \bm{W} = \begin{bmatrix}
        0 & -0.0910 & 0.0219 \\
        0.2472 & 0 & -0.2203 \\
        0.1194 & 0.4430 & 0
    \end{bmatrix}
    \; .
\end{equation}
In temperature range $1.70899 < T < 1.98262$ there are two degenerate ideal-gas matrices. For example, at $T = 1.8$ they are
\begin{equation}
    \bm{W}^{(1)} = \begin{bmatrix}
        0 & -0.1556 & 0.2948 \\
        0.3972 & 0 & -0.2540 \\
        -0.0950 & 0.5695 & 0
    \end{bmatrix}
    \; ,
    \quad 
       \bm{W}^{(2)} = \begin{bmatrix}
        0 & -0.2680 & -0.1982 \\
        0.3792 & 0 & -0.2800\\
        0.4164 & 0.3999 & 0
    \end{bmatrix}
    \; .
\end{equation}
Notice that $\bm{W}^{(1)}$ is cyclic-dominant, and $\bm{W}^{(2)}$ contains a fully inhibitory unit, a fully excitatory unit, and a mixed unit.

In the temperature range $1.42383 < T < 1.70899$ there are four degenerate ideal-gas matrices. For example at $T = 1.6$ they are:
\begin{eqnarray}
& \bm{W}^{(1)}  =
\begin{bmatrix}
 0      & -0.2267  &  0.4457  \\
 0.5416 &  0       & -0.2859 \\
-0.1648 &  0.6876  &  0
\end{bmatrix}
\; ,
 \quad 
& \bm{W}^{(2)}  =
\begin{bmatrix}
0      & -0.4066  & -0.2910 \\
0.4855 &  0       & -0.3733\\
0.5803 &  0.4041  &  0
\end{bmatrix}
\; ,
\\
& \bm{W}^{(3)}  =
\begin{bmatrix}
0       &  0.3287  & -0.3768  \\
0.0872  &  0       &  0.6061 \\
0.6159  & -0.3473  &  0
    \end{bmatrix}
    \; ,
  \quad 
& \bm{W}^{(4)}  = \begin{bmatrix}
0      & -0.0565 & -0.4968 \\
0.4502 &  0      &  0.4151 \\
0.6007 & -0.3730 & 0
    \end{bmatrix}
    \; .
\end{eqnarray}
Matrices $\bm{W}^{(1)}$ and $\bm{W}^{(3)}$ are cyclic-dominant, and $\bm{W}^{(2)}$ and $\bm{W}^{(4)}$ contain a fully inhibitory unit, a fully excitatory unit, and a mixed unit. 

In the temperature range $1.1927 < T < 1.42383$ there are six degenerate ideal-gas matrices. For example at $T = 1.3$ they are:
\begin{eqnarray}
 & \bm{W}^{(1)}  =
\begin{bmatrix}
 0      & -0.3138  &  0.6633  \\
 0.7637 &  0       & -0.3303 \\
-0.2300 &  0.8906  &  0
\end{bmatrix}
\; , 
 \quad 
& \bm{W}^{(2)}  =
\begin{bmatrix}
0      & -0.6144  & -0.4012 \\
0.6246 &  0       & -0.5497\\
0.8160 &  0.4247  &  0
\end{bmatrix}
\; , \\
& \bm{W}^{(3)}  =
\begin{bmatrix}
0       &  0.6076  & -0.4114  \\
-0.0348  &  0       &  0.8313 \\
0.8193  & -0.4182  &  0
    \end{bmatrix}
\; ,
\quad 
& \bm{W}^{(4)}  = \begin{bmatrix}
0      & -0.2165 & -0.7011 \\
0.7288 &  0      &  0.4015 \\
0.7353 & -0.5527 & 0
    \end{bmatrix}
    \; , \\
& \bm{W}^{(5)}  =
\begin{bmatrix}
0        &  0.7334  & -0.0233  \\
-0.5178  &  0       & -0.6513 \\
-0.4222  &  0.8172  &  0
    \end{bmatrix}
\; ,
\quad 
& \bm{W}^{(6)}  = \begin{bmatrix}
0       & 0.6304  & -0.3756 \\
-0.6115 & 0       & -0.5642 \\
-0.0495 & 0.9185  & 0
    \end{bmatrix}
    \; .
\end{eqnarray}
Matrices $\bm{W}^{(1)}$ and $\bm{W}^{(3)}$ are cyclic-dominant, and $\bm{W}^{(2)}$ and $\bm{W}^{(4)}$ contain a fully inhibitory unit, a fully excitatory unit, and a mixed unit, and matrices $\bm{W}^{(5)}$ and $\bm{W}^{(6)}$ are two excitation-inhibition balanced network with one inhibitory unit and two excitatory units.

In the temperature range $0.31636 < T < 1.1927$ there are eight degenerate ideal-gas matrices. For example at $T = 0.7$ they are:
\begin{eqnarray}
 & \bm{W}^{(1)}  =
\begin{bmatrix}
 0      & -0.4600  &  1.2828  \\
 1.4070 &  0       & -0.4041 \\
-0.2930 &  1.5306  &  0
\end{bmatrix}
\; , 
\quad 
& \bm{W}^{(2)}  =
\begin{bmatrix}
0      & -1.2095  & -0.6280 \\
0.9262 &  0       & -1.1336 \\
1.4851 &  0.4724  &  0
\end{bmatrix}
\; , 
\\
& \bm{W}^{(3)}  =
\begin{bmatrix}
0       &  1.2734  & -0.4853  \\
-0.1377 &  0       &  1.4574 \\
1.4611  & -0.5419  &  0
\end{bmatrix}
\; ,
\quad 
& \bm{W}^{(4)}  = 
\begin{bmatrix}
0      & -0.4555 & -1.2844 \\
1.4063 &  0      &  0.4066 \\
1.0425 & -1.1583 &  0
    \end{bmatrix}
    \; , \\
& \bm{W}^{(5)}  =
\begin{bmatrix}
0        &  1.3581  &  0.1128  \\
-0.7905  &  0       & -1.2321 \\
-1.1914  &  1.0046  &  0
    \end{bmatrix}
\; ,
\quad 
& \bm{W}^{(6)} =
\begin{bmatrix}
0       & 0.8381  & -1.0746 \\
-1.2420 & 0       & -0.7747 \\
 0.0551 & 1.5574  & 0
\end{bmatrix}
\; , \\
& \bm{W}^{(7)}  =
\begin{bmatrix}
0        &  0.1252  &  1.3570  \\
-1.0628  &  0       &  1.0067 \\
-0.9131  &  -1.2628  &  0
\end{bmatrix}
\; ,
\quad 
& \bm{W}^{(8)} =
\begin{bmatrix}
0       & -0.9769  & 0.9502 \\
0.0145  &  0       & 1.4638 \\
-1.3771 & -0.7295  & 0
\end{bmatrix}
\; .
\end{eqnarray}
Matrices $\bm{W}^{(1)}$ and $\bm{W}^{(3)}$ are cyclic-dominant, and the other six matrices are all formed by three functionally distinct units.

The degeneracy then decreases at lower temperatures. First the matrix $\bm{W}^{(1)}$ attaches the eigenvalue bottomline at $T = 0.31636$, followed by matrix $\bm{W}^{(3)}$ at $T = 0.31346$, followed by $\bm{W}^{(6)}$ at $T = 0.23364$, followed by $\bm{W}^{(8)}$ at $T = 0.22835$, followed by $\bm{W}^{(5)}$ at $T = 0.21387$,  followed by $\bm{W}^{(4)}$ at $T = 0.21290$,  followed by $\bm{W}^{(2)}$ at $T = 0.19830$, and finally followed by $\bm{W}^{(7)}$ at $T = 0.19402$. At the lowest temperature $T = 0.19402$ the ideal-gas law matrix is
\begin{equation}
    \bm{W}^{(7)} = 
    \begin{bmatrix}
        0 & 0.1833 & 3.0454 \\
        -2.8955 & 0 & 1.3983 \\
        -1.4573 & -3.0408 & 0
    \end{bmatrix}
    \; .
\end{equation}

The above detailed results on the problem instance (\ref{eq:Cgeneral}) demonstrate that the abundance of ideal-gas optimal matrices. These results indicate that, when the temperature $T$ is in an appropriate range, the ideal-gas law optimal solutions might be relatively easy to obtain. We expect these aspects also extend to higher-dimensional systems with $N > 3$ units.

\subsubsection*{Phase diagram for uniform input correlation matrix}

For the $N=3$ system with symmetric sensory inputs the phase diagram in the plane of $c$ and $T$ is shown in Fig.~\ref{fig:PDN3}. This phase diagram is qualitatively similar to the one for $N=5$ (Fig.~3 of the main text) and it is much more richer than the phase diagram for $N=2$ (Fig.~1 of the main text), confirming the intuitive expectation that three-body and multi-body (effective) interactions are capable of producing high degrees of complexity. 

\begin{figure}[h]
  \centering
  \includegraphics[angle=270,width=0.5\textwidth]{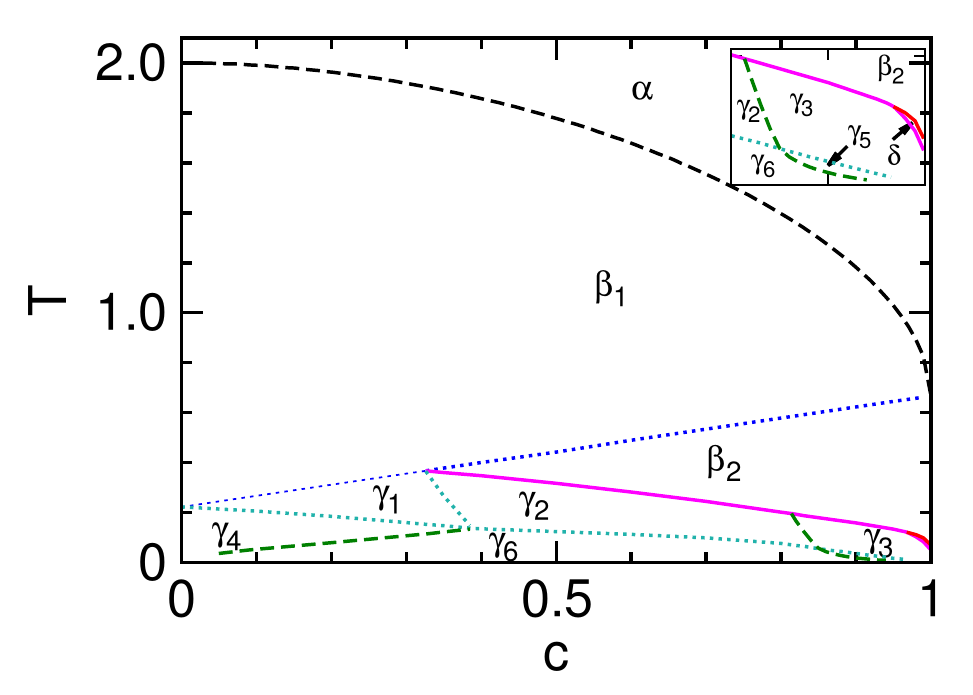}
  \caption{
    Phase diagram for the $N=3$ system with symmetric inputs ($c_{i i} = 1$ and pair correlation $c_{i j} = c$ for any $i \neq j$). Region $\alpha$ is the permutation-symmetric phase; region $\beta_1$: one-component and two-component optimal networks obeying the ideal-gas law; region $\beta_2$: one-component networks with cyclic dominance; region $\gamma_1$: two-component networks obeying the ideal-gas law; region $\gamma_2$--$\gamma_6$: two-component networks with excitation-inhibition balance; region $\delta$ is an intermediate region between $\beta_2$ and $\gamma_3$. Dashed lines indicate continuous transitions associated with symmetry change; dotted lines indicate continuous transitions caused by the eigenvalue bottomline constraint; solid lines mark discontinuous transition between different types of networks.}
  \label{fig:PDN3}
\end{figure}

We briefly describe the properties of these different phase space regions here.
\begin{enumerate}
\item[$\alpha$:] one-component networks with permutation symmetry.
\item[$\beta_1$:] containing both one-component networks with cyclic dominance and two-component networks with excitation-inhibition balance. The ideal-gas law is obeyed.
\item[$\beta_2$:] one-component networks with cyclic dominance, constrained by eigenvalue bottomline.
\item[$\gamma_1$:] two-component EI-balanced networks without internal permutation symmetry within the group of two units. The ideal-gas law is obeyed. One example network obtained at $c=0.35$ and $T=0.2643$ (slightly below the $\gamma_2 \rightarrow \gamma_1$ critical temperature value $0.2644$) is
\begin{equation}
  \gamma_1: \quad 
  \begin{bmatrix}
    0       & -1.7971  & -1.8265 \\
    2.2684  & 0        & -1.1918 \\
    2.2407  & -1.2430  & 0
  \end{bmatrix}
  \; ,
\end{equation}
whose entropy is $S = -2.87$ and energy $E = 0.3965$, and the eigenvalues of $\bm{I}+\bm{W}$ are: $\{ 2.2177, \; 0.3912 \pm 2.7928i \}$.
\item[$\gamma_2$:] two-component EI-balanced networks with internal permutation symmetry within the group of two units. One example for $c=0.35$ and $T = 0.2661$ (slightly above the $\gamma_2\rightarrow \gamma_1$ critical temperature $0.2644$) is
  \begin{equation}
    \gamma_2 : \quad
    \begin{bmatrix}
      0       & -1.8045  & -1.8044 \\
      2.2471  & 0        & -1.2107 \\
      2.2470  & -1.2105  & 0
    \end{bmatrix}
    \; ,
  \end{equation}
  whose entropy $S = -2.86$ and energy $E = 0.3992$) and the eigenvalues of $\bm{I}+\bm{W}$ are  $\{ 2.2106, \;  0.3947 \pm 2.7827 i \}$.
\item[$\gamma_3$:] two-component EI-balanced networks at high values of $c$ without permutation symmetry within the group of two units.
\item[$\gamma_4$:] two-component EI-balanced networks at low values of $c$ constrained by the eigenvalue bottomline. One example obtained at $c=0.35$ and $T = 0.1404$ is
  \begin{equation}
    \gamma_4: \quad 
    \begin{bmatrix}
      0      & -2.1783 & -2.9210 \\
      3.3757 &  0      & -1.2982 \\
      2.6520 & -2.4645 &  0
    \end{bmatrix}
    \; ,
  \end{equation}
  whose entropy $S = -3.8$ and energy $E = 0.21335$, and the eigenvalues of $\bm{I}+\bm{W}$ are $\{ 2.99998,\; 10^{-5} \pm 3.8601 i\}$. There is no permutation symmetry within the group of two units.
\item[$\gamma_5$:] two-component EI-balanced networks at high values of $c$ constrained by the eigenvalue bottomline.  There is no permutation symmetry within the group of two units.
\item[$\gamma_6$:] two-component EI-balanced networks constrained by the eigenvalue bottomline, with permutation symmetry within the group of two units. The transition between $\gamma_4$ and $\gamma_6$ and that between $\gamma_5$ and $\gamma_6$ are both continuous. An example obtained at $c=0.35$ and $T=0.0464$ is
  \begin{equation}
    \gamma_6: \quad
    \begin{bmatrix}
      0      &  -4.4268 & -4.4269 \\
      5.1689 &  0       & -1.9998 \\
      5.1689 &  -2.0001 & 0
    \end{bmatrix}
    \; ,
  \end{equation}
  whose entropy $S= -4.9$ and $E=0.1198$ and the eigenvalues of $\bm{I}+\bm{W}$ are $\{ 2.99998 ,\; 10^{-5} \pm 6.6906 i\}$. The permutation within the two excitatory units $2$ and $3$ is preserved in phase region $\gamma_6$.
\item[$\delta$:] An intermediate phase region between $\beta_2$ and $\gamma_3$ at high values of $c$.
\end{enumerate}

Our numerical computations suggest that the intermediate phase region $\delta$ only exists for $c > 0.9678$. To demonstrate the discontinuous transitions between $\beta_2$ and $\delta$ and between $\delta$ and $\gamma_3$, we plot the free energy curves of these three phases for $c=0.99$ at low temperatures $T$ close to $0.09$ in Fig.~\ref{fig:N3a0p99A}.

\begin{figure}[h]
  \centering
  \includegraphics[angle=270,width=0.5\textwidth]{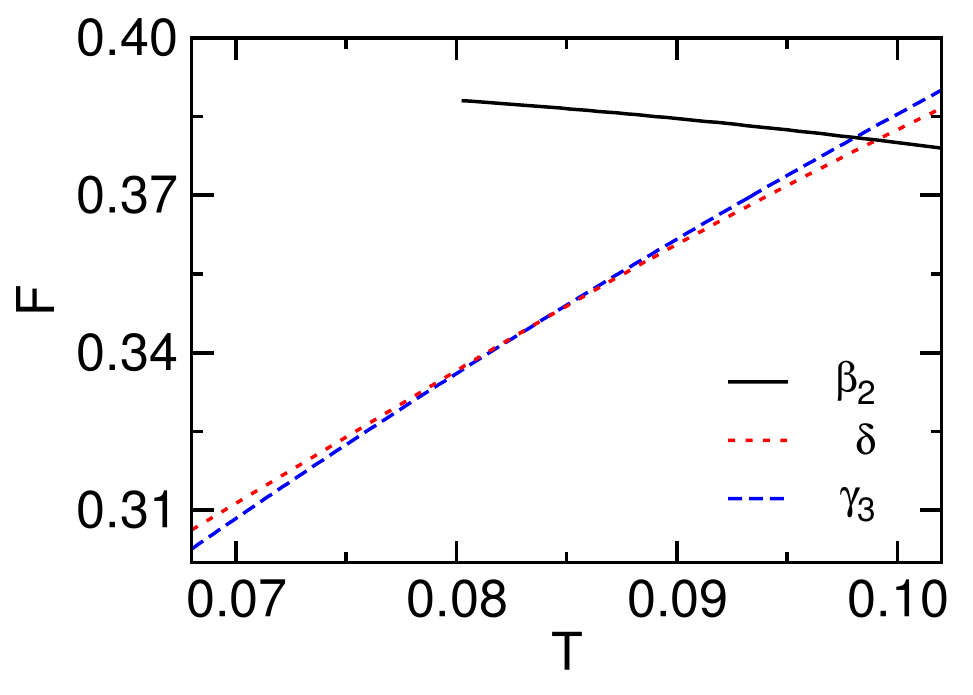}
  \caption{
    Free energy of the three phases ($\beta_2$, $\delta$, and $\gamma_3$) obtained for $N=3$ with input correlation $c=0.99$ at low temperatures $T$. 
  }
  \label{fig:N3a0p99A}
\end{figure}

There is a discontinuous transition between the $\beta_2$ phase and the $\delta$ phase at $T = 0.09905$, at which the weight matrix changes from a purely inhibitory one with $S = 0.5$ and $E = 0.4300$ (the eigenvalues of $\bm{I}+\bm{W}$ are $\{ 2.99998, \; 10^{-5} \pm 0.4496 i \}$) to one with a single strong excitatory weight and five inhibitory weights whose  $S = -2.1$ and $E = 0.1725$ (the eigenvalues of $\bm{I}+\bm{W}$ are $\{ 2.99998, \; 10^{-5} \pm 1.6499 i \}$:
\begin{equation}
  \beta_2: \quad
  \begin{bmatrix}
    0 & 1.2596 & 0.7404 \\
    0.7404 & 0 & 1.2596 \\
    1.2596 & 0.7404 & 0
  \end{bmatrix}
  \; ,
  \quad \quad
  \delta: \quad
  \begin{bmatrix}
    0 & 4.1525 & 0.2186 \\
    -2.1201 & 0 & 2.5863 \\
    0.8416 & 3.4403 & 0
  \end{bmatrix}
  \; .
\end{equation}
This phase transition brings a large drop in the energy and in the entropy. At a lower temperature $T=0.08369$, there is another discontinuous transition, from a matrix of the $\delta$ phase with $S = -2.45$ and $E = 0.1407$ (the eigenvalues of $\bm{I}+\bm{W}$ are $\{ 2.99998, \; 10^{-5} \pm 1.9645 i \}$) to an EI-balanced matrix of the $\gamma_3$ phase with $S = -2.6$ and $E = 0.1281$ (the eigen values of $\bm{I}+\bm{W}$ are $\{  0.6148, \; 1.1926\pm 4.5251 i\}$):
\begin{equation}
  \delta : \quad
  \begin{bmatrix}
    0 & 4.7250 & 0.1301 \\
    -2.5814 & 0 & 2.8358 \\
    0.8251 & 3.9590 & 0
  \end{bmatrix}
  \; ,
  \quad \quad
  \gamma_3: \quad
  \begin{bmatrix}
    0       & 5.1667 & 0.3121 \\
    -4.1575 & 0      & 0.5176 \\
    -1.4773 & 3.0449 & 0
  \end{bmatrix}
\; .
\end{equation}
This second discontinuous transition only causes a relatively small drop in the energy and in the entropy, but the change in the optimal matrix $\bm{W}$ is quite comprehensive.

\subsection{Some numerical results for $N = 10$ and $c = 0.8$}

Besides $N=3$ and $N=5$, we also study systems with other number of units up to $N=10$. The phase diagram for the system with $N = 4$ units has also been obtained by numerical computations. It is qualitatively similar to the phase diagrams of $N=3$ and $N=5$. Here we show some numerical results obtained on the large system of $N=10$, which reveal the competition among multiple local minima of the free energy. The input pair correlation is fixed to be $c=0.8$ for simplicity. 

We confirm that there is a continuous permutation-symmetry-breaking transition at a high critical temperature $T_c^{(1)} = 0.9489$ for $N=10$ and $c=0.8$, at which cyclic dominance starts to emerge, and the energy versus temperature relationship starts to follow the ideal-gas law $E = 5 T$. The optimal solutions are not unique but are degenerate. For example, we observe two distinct solutions $\bm{W}$ with the same entropy $S = -4.6$ and the same mean energy $E = 1.1555$ at temperature $T = 0.2311$. One of them is one-component with mostly positive synaptic weights, while the other is a two-component EI-balanced solution containing a single purely excitatory unit and nine inhibitory units.

As temperature decreases to another critical value $T_c^{(2)} = 0.1763$, the ideal-gas law $E = 5 T$ finally breaks as all the complex eigenvalues of the optimal solutions reach the bottomline. The energy susceptibility then drops from $5$ to $4.5$. At temperatures slightly below this critical value the minimum free energy solution is the two-component EI-balanced one with group $g_E$ containing a single excitatory unit and group $g_I$ containing the remaining nine inhibitory units (we call this the $1e$--$9i$ type structure).

As temperature decreases to much below $T_c^{(2)}$, we find that two other types of EI-balanced networks, the $3e$--$7i$ type with $g_E$ containing three excitatory units and $g_I$ containing seven inhibitory units and the $5e$--$5i$ type with $g_E$ containing five excitatory units and $g_I$ containing five inhibitory units, start to compete with this $1e$--$9i$ type networks. To demonstrate this competition, we draw in Fig.~\ref{fig:FEN10a0p8T0p107} the free energy differences of these three types of EI-balanced solutions at the vicinity of $T = 0.107$.

\begin{figure}[h]
  \centering
  \subfigure[]{
    \includegraphics[angle=270,width=0.45\textwidth]{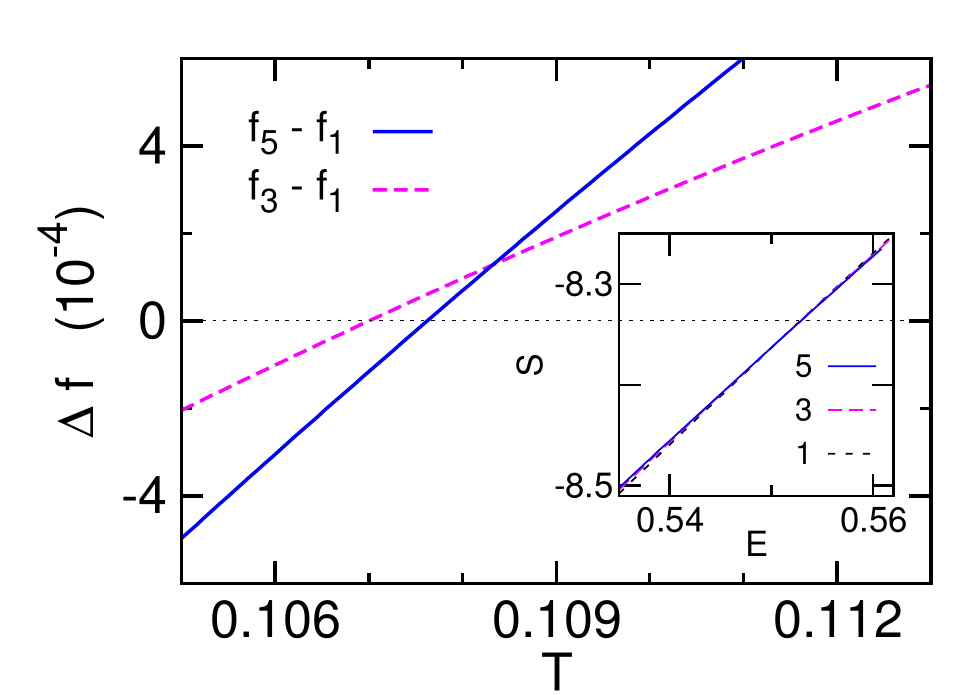}
    \label{fig:FEN10a0p8T0p107}
  }
  \subfigure[]{
    \includegraphics[angle=270,width=0.45\textwidth]{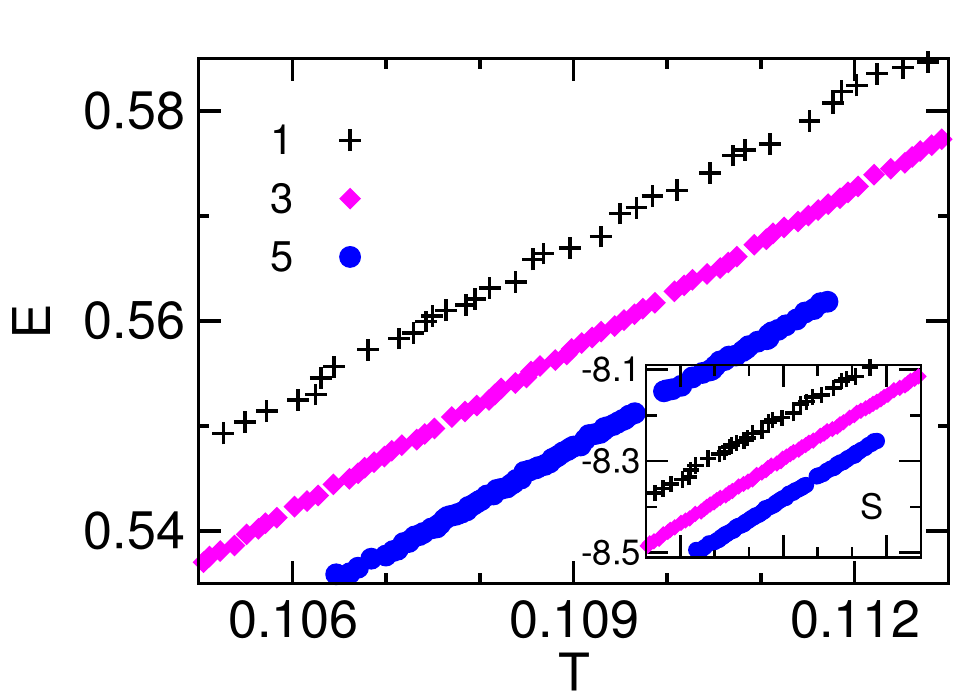}
    \label{fig:EnN10a0p8T0p107}
  }
  \caption{
    Results obtained for $N=10$, $c=0.8$, indicating a discontinuous phase transition at $T = 0.1077$. (a) Free energy difference. The $1e$--$9i$, $3e$--$7i$ and $5e$--$5i$ type solutions are indicated by the labels $1$, $3$ and $5$, respectively, and their free energies are $f_1$, $f_3$ and $f_5$. The inset shows the entropy-energy curves for the three types of solutions. (b) Energy versus temperature, and entropy versus temperature (inset).
  }
  \label{fig:N10a0p8T0p107}
\end{figure}

The free energy $f_5$ of the $5e$--$5i$ solution starts to be lower than the value $f_3$ of the $3e$--$7i$ solution as temperature is decreased below the threshold value $0.1084$, so the $5e$--$5i$ structure becomes more favorable than the $3e$--$7i$ structure. At the lower critical value $0.1077$ the $5e$--$5i$ structure also starts to be dominating over the $1e$--$9i$ structure. The $3e$--$7i$ structure also has lower free energy than the $1e$--$9i$ structure at $T < 0.1070$, but it can not compete with the $5e$--$5i$ structure. We therefore conclude that there is a discontinuous phase transition at $T_c^{(3)} = 0.1077$ between the $1e$--$9i$ and $5e$--$5i$ EI-balanced network structures. This transition is associated with a discontinuity in the entropy and in the mean energy [Fig.~\ref{fig:EnN10a0p8T0p107}].

To have a concrete feeling of the three types of EI-balanced network structures, we list three example matrices here. First is the $1e$--$9i$ network with entropy $S = -8.35$:  
\begin{equation}
\begin{bmatrix}
    0      & 1.197 & 1.195  & 1.193  & 1.206  & 1.214 & 1.200 & 1.216 & 1.202 & 1.191 \\
    -1.079 & 0     & -0.358 & 1.746  & 1.442  & 1.264 & 0.252 & 1.290 & 1.273 & 1.080 \\
    -1.072 & 2.357 & 0      & 0.493  & 1.531  & 0.077 & 0.942 & 1.101 & 0.669 & 0.837 \\
    -1.067 & 0.251 & 1.499  & 0      & 1.503  & 0.439 & 0.650 & 0.547 & 0.848 & 2.283 \\
    -1.088 & 0.561 & 0.468  & 0.494  & 0      & 1.375 & 2.101 & 1.667 & 0.107 & 1.208 \\
    -1.085 & 0.743 & 1.925  & 1.562  & 0.631  & 0     & 0.326 & 1.940 & 0.468 & 0.393 \\
    -1.069 & 1.742 & 1.050  & 1.340  & -0.104 & 1.674 & 0     & 0.178 & 0.656 & 1.479 \\
    -1.080 & 0.714 & 0.901  & 1.450  & 0.339  & 0.067 & 1.821 & 0     & 1.992 & 0.730 \\
    -1.081 & 0.730 & 1.327  & 1.147  & 1.896  & 1.537 & 1.339 & 0.009 & 0     & 0.010 \\
    -1.080 & 0.921 & 1.163  & -0.286 & 0.793  & 1.613 & 0.516 & 1.273 & 1.990 & 0
\end{bmatrix}
\; ,
\end{equation}
whose energy is $E = 0.55036$ and the eigenvalues of $\bm{I}+\bm{W}$ are $\{ 7.083736231 , \; 2.916183746 , \; 10^{-5} \pm 1.9483 i , \; 10^{-5} \pm 1.9477 i , \; 10^{-5} \pm 1.9472 i , \; 10^{-5} \pm 1.9464 i\}$. The $3e$-$7i$ network with entropy $S = -8.35$ is:
\begin{equation}
  \begin{bmatrix}
    0      & -1.020 & -1.020 & 1.249 & 1.250  & 1.248  & 1.245  & 1.246 & 1.244  & 1.246 \\
    -1.020 & 0      & -1.020 & 1.249 & 1.250  & 1.248  & 1.245  & 1.246 & 1.244  & 1.246 \\
    -1.020 & -1.020 & 0     & 1.249 & 1.250  & 1.248  & 1.245  & 1.246 & 1.244  & 1.246 \\
    -1.060 & -1.060 & -1.060 & 0     & 1.512  & 1.660  & 1.730  & 1.062 & -0.067 & 0.099 \\
    -1.059 & -1.059 & -1.059 & 0.489 & 0      & 0.380  & 1.951  & 0.295 & 0.837  & 2.050 \\
    -1.059 & -1.059 & -1.059 & 0.342 & 1.624  & 0      & 0.848  & 0.633 & 2.289  & 0.268 \\
    -1.057 & -1.057 & -1.057 & 0.272 & 0.050  & 1.152  & 0      & 2.302 & 1.113  & 1.118 \\
    -1.060 & -1.060 & -1.060 & 0.938 & 1.705  & 1.368  & -0.305 & 0     & 0.506  & 1.783 \\
    -1.059 & -1.059 & -1.059 & 2.066 & 1.166  & -0.290 & 0.887  & 1.491 & 0      & 0.680 \\
    -1.059 & -1.059 & -1.059 & 1.901 & -0.051 & 1.733  & 0.880  & 0.218 & 1.319  & 0
  \end{bmatrix}
  \; ,
\end{equation}
whose energy is $E = 0.55031$ and the corresponding eigenvalues are $\{ 2.0196, \; 2.0196, \; 2.9803 \pm 3.4025 i, \; 10^{-5} \pm 1.9277 i , \; 10^{-5} \pm 1.9268 i , \; 10^{-5} \pm 1.9265 i\}$. There is permutation symmetry within the three units of group $g_E$. The $5e$--$5i$ network with $S = -8.36$ is
\begin{equation}
  \begin{bmatrix}
    0     & -0.994 & -0.994 & -0.994 & -0.994 & 1.337 &  1.330 &  1.334 &  1.339 &  1.336 \\
    -0.994 & 0      & -0.994 & -0.994 & -0.994 & 1.337 &  1.330 &  1.334 &  1.339 &  1.336 \\
    -0.994 & -0.994 &  0     & -0.994 & -0.994 & 1.337 &  1.330 &  1.334 &  1.339 &  1.336 \\
    -0.994 & -0.994 & -0.994 &  0     & -0.994 & 1.337 &  1.330 &  1.334 &  1.339 &  1.336 \\
    -0.994 & -0.994 & -0.994 & -0.994 &  0     & 1.337 &  1.330 &  1.334 &  1.339 &  1.336 \\
    -1.054 & -1.054 & -1.054 & -1.054 & -1.054 & 0     & -0.483 &  1.553 &  1.410 &  1.508 \\
    -1.051 & -1.051 & -1.051 & -1.051 & -1.051 & 2.481 &  0    &  0.435 &  0.501 &  0.586 \\
    -1.051 & -1.051 & -1.051 & -1.051 & -1.051 & 0.448 &  1.562 &  0    & -0.071 &  2.071 \\
    -1.053 & -1.053 & -1.053 & -1.053 & -1.053 & 0.591 &  1.499 &  2.072 &  0     & -0.160 \\
    -1.053 & -1.053 & -1.053 & -1.053 & -1.053 & 0.494 &  1.413 & -0.072 &  2.161 &  0
  \end{bmatrix}
\; ,
\end{equation}
whose energy is $E = 0.55032$ and the eigenvalues of $\bm{I}+\bm{W}$ are $\{ 1.9942 \pm 2 \times 10^{-6} i , \; 1.9942 \pm 2 \times 10^{-5} , \; 1.0115 \pm 4.3843 i  , \; 10^{-5} \pm 1.9117 i , \; 10^{-5} \pm 1.9108 i \}$. There is permutation symmetry within the five units of group $g_E$ and cyclic dominance within the five units of group $g_I$.

\subsection{The effect of restricting all the synaptic weights to be non-negative}

 Some neurons in the biological brain are purely inhibitory, and the networks formed by them will have no internal excitations. If the predictive coding system is composed purely of inhibitory units, then all the synaptic weights will be non-negative. Figure~\ref{fig:N2EW:nnw} has clearly demonstrated the significant effect at low temperatures of restricting the synaptic weights to be non-negative. Here we present some additional results obtained on multi-unit systems ($N=5$). To compare with Fig.~2 of the main text, we fix the input correlation to be $c = 0.8$, see Fig.~\ref{fig:N5EWPW}.

\begin{figure}[t]
  \centering
  \includegraphics[angle=270,width=1.0\linewidth]{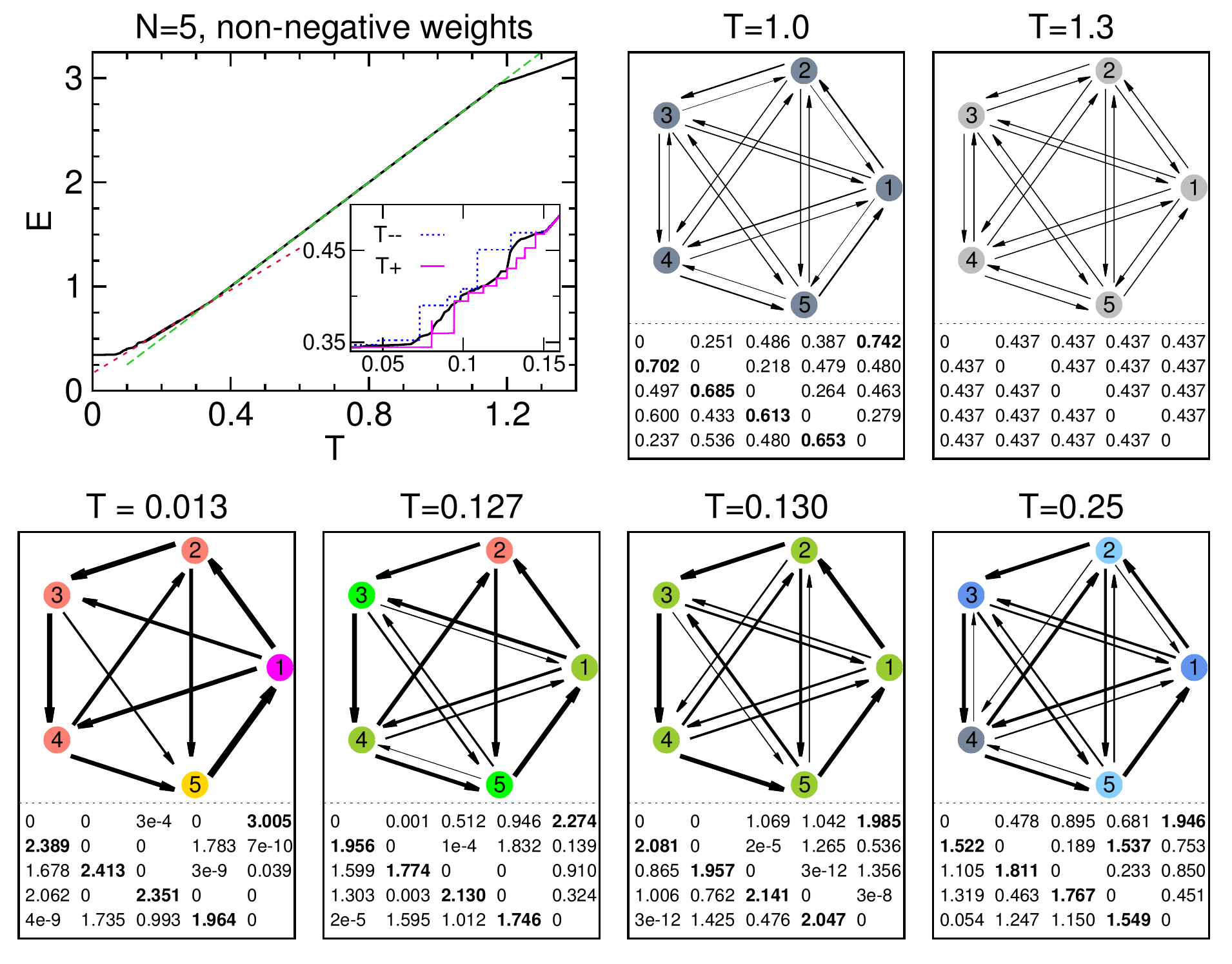}
  \caption{
    Optimal predictive coding with non-negative weights for $N=5$ and input pair correlation $c=0.8$. The energy $E$ has two kinks at $T=1.179$ and $T=0.336$ (the slopes of fitting lines are respectively $2.5$ and $2.0$), and it experiences a series of small drops at low temperatures $T< 0.15$. The inset demonstrates the energy hysteresis during a slow cycle of $T$ increasing (solid thin line) and decreasing (dotted thin line) in comparison with equilibrium (thick solid line). Also shown are the optimal weight matrices at several $T$ values and the corresponding interaction graphs. The units are clustered into groups and are distinguished using different colors in each graph. The width of link from unit $j$ to unit $i$ is proportional to the weight $w_{i j}$ (if $w_{i j}$ is less than $10\%$ of the maximum value, the edge is not shown). 
  }
  \label{fig:N5EWPW}
\end{figure}

If the temperature is high enough the restriction of all weights being non-negative has no effect to the system's property. The weight matrix is completely symmetric at high temperatures  ($w_{i j} = w_{j i} = w > 0$). A continuous transition occurs at $T=1.179$ with reciprocity breaking ($w_{i j} \neq w_{j i}$) and with the formation of cyclic dominance among any set of three units and the emergence of one global strongest inhibition cycle involving all the units. The energy susceptibility, $\chi \equiv {\rm d} E/ {\rm d} T$, has a discontinuity at this transition point, and the ideal-gas law follows after this permutation-symmetry breaking transition.

As $T$ decreases to $0.336$ the susceptibility drops from $2.5$ to $2.0$ but the energy $E$ itself is continuous, indicating another continuous transition. Another signature of this transition is that the real parts of all the complex eigenvalues of $\bm{I}+\bm{W}$ touch the bottomline value $10^{-5}$. At this transition point all the synaptic weights are still positive, so the non-negative constraint does not affect this transition.

As $T$ further decreases to $0.15$ some of the weights reach zero and the mean energy $E$ starts to level off. There is then a considerable drop in $E$ at $T =  0.128$ which is caused by a sudden relatively huge change in the weight matrix (see Fig.~\ref{fig:N5EWPW}, $T=0.127$ and $0.130$).

As $T$ keeps decreasing, the weight matrix experiences several additional abrupt changes before reaching the completely one-directional form at very low $T$ values. The discontinuous nature of weights change is confirmed by the strong hysteresis behavior of energy at low temperatures [Fig.~\ref{fig:N5EWPW}, inset]. The nearly minimum-energy weight matrix at $T=0.013$ has a clear hierarchical structure: neuron $1$ is inhibited by neuron $5$ and inhibits neuron group $\{2,3,4\}$ which in turn inhibits neuron $5$, and there are internal cyclic inhibitions within group $\{2,3,4\}$ and within group $\{1, 2, 5\}$. Because excitation--inhibition balance can not be realized in the system, at low temperatures the ideal-gas law has no chance of being approximately restored. The mean energy saturates to a value slightly below $0.35$ as $T$ decreases to zero.

\clearpage

\end{widetext}

\end{document}